\documentclass[notitlepage,onecolumn,secnumarabic,amssymb, nobibnotes, aps,nofootinbib, prd]{revtex4}
\usepackage{graphicx,epsfig}
\usepackage{natbib}
\usepackage{upgreek}
\usepackage{bm}
\usepackage{amsmath}
\usepackage{amsfonts}
\usepackage{xcolor}
\usepackage{journal_names}
\usepackage{hyperref}
\usepackage{relsize,esint}
\usepackage{wrapfig}
\usepackage{color}

\newcounter{box}
\newcounter{temp}

\usepackage{amsmath}

\newenvironment{boxequation}{%
  \setcounter{temp}{\value{equation}}%
  \setcounter{equation}{\value{box}}%
}{%
  \setcounter{box}{\value{equation}}%
  \setcounter{equation}{\value{temp}}%
}
\newcommand{\bb}{\begin{boxequation}}
\newcommand{\eb}{\end{boxequation}}

\def\d {\mbox{d}}

\def\be{\begin{equation}}
\def\ee{\end{equation}}
\def\bea{\begin{eqnarray}}
\def\eea{\end{eqnarray}}

\def\n {\bm{n}}

\def\x {\bm{x}}

\newcommand{\HH}{\mathcal{H}}
\newcommand{\DD} {\mbox{D}}
\newcommand{\D}{\nabla}

\renewcommand{\>}{\rangle}

\newcommand{\two}{^{\text{\tiny (2)}}}

\newcommand{\twod}{_{\text{\tiny (2)}}}

\def\S{^{\text{\tiny S}}}
\def\V{^{\text{\tiny V}}}
\def\T{^{\text{\tiny T}}}

\def\Sd{_{\text{\tiny S}}}
\def\Vd{_{\text{\tiny V}}}
\def\Td{_{\text{\tiny T}}}
\def\NLd{_{\text{\tiny Q}}}
\def\Q{_{\text{\tiny Q}}}

\newcommand{\I}{{\int_{{\!}_\lambda}}}

\newcommand{\p}{_{{\text{\tiny$\|$}}}}
\newcommand{\pp}{{{\text{\tiny$\|$}}}}
\renewcommand{\o}{{\hspace{-0.5mm}{\text{\tiny$\perp$}}}}
\newcommand{\po}{{\hspace{-0.5mm}{\text{\tiny$\perp\!\!\!|$}}}}
\renewcommand{\bot}{\o}

\newcommand{\tot}[1]{#1}
\newcommand{\back}[1]{\bar{#1}}

\newcommand{\lambdaz}{\nu}
\newcommand{\Iz}{{\int_{{\!}_\lambdaz}}}

\newcommand{\Is}{{\int_{{\!}_{\lambda_s}}}}
\newcommand{\Ic}{{\int_{{\!}_{\tiny\chi}}}}

\newcommand{\Icpr}{{\int_{{\!}_{\tilde\chi}}}}
\newcommand{\Ipr}{{\int_{{\!}_{\tilde\lambda}}}}

\renewcommand{\tt}[1]{\bf\emph{#1}}

\begin{document}

\title{Nonlinear relativistic corrections to cosmological distances,\\redshift and gravitational lensing magnification. II -- Derivation}

\author{ Obinna Umeh$^{1,2}$, Chris Clarkson$^2$ and  Roy Maartens$^{1,3}$,\\
\emph{$^1$Physics Department, University of the Western Cape,
Cape Town 7535, South Africa \\
$^2$Astrophysics, Cosmology \& Gravity Centre, and, Department of Mathematics \& Applied Mathematics, University of Cape Town, Rondebosch 7701, South Africa\\
$^3$Institute of Cosmology \& Gravitation, University of Portsmouth, Portsmouth PO1 3FX, United Kingdom\\
}}
\date{\today}

\begin{abstract}

We present a derivation of the cosmological distance-redshift relation up to second order in perturbation theory. In addition, we find the observed redshift and the lensing magnification to second order.
We do not require that the density contrast is small, we only that the metric potentials and peculiar velocities are small. Thus our results apply into the nonlinear regime, and can be used for most dark energy models. We present the results in a form which can be readily computed in an N-body simulation. This paper accompanies Paper I, where the key results are summarised in a physically transparent form and applications are discussed.


\end{abstract}

\maketitle
 \setcounter{footnote}{0}
\DeclareGraphicsRule{.wmf}{bmp}{jpg}{}{}
\maketitle

\section{Introduction}

The next generation of cosmological surveys will map the universe to extremely high precision across a vast range of scales. These surveys may be sensitive to a variety of subtle nonlinear relativistic effects~-- most of which remain to be analyzed. The angular diameter-redshift relation, which is a key determinant of any cosmological model, is affected by inhomogeneities along the line of sight. Nonlinear contributions change apparent sizes and brightnesses of objects, beyond the usual convergence contributions from standard gravitational lensing~\cite{Bartelmann:1999yn} and from the relativistic linear correction in the form of Doppler lensing~\cite{Bonvin:2008ni,Bacon:2014uja}. These corrections could well be significant: for example, a deep spherical void can change the magnitude-redshift relation by up to 20\% beyond the prediction of linear perturbation theory when all relativistic effects are taken into account~\cite{Bolejko:2012uj}. Accurately quantifying the amplitude of the corrections around 
realistic 
structures is important for the precise modelling now required for large-scale surveys. 

Light  propagation  in inhomogeneous  spacetimes  gives  rise to
modification of the area distance  due to both
Ricci focussing (from the matter along the ray) and Weyl
focussing (from the tidal effects of nearby matter). It leads to corrections in redshift due to the differences between the true (observed) redshift of a source
and its redshift in an average smoothed-out model. 
For light propagating in a perturbed Friedmann-Lemaitre-Robertson-Walker (FLRW)
space-time, linear perturbations of the angular diameter, or area, distance were computed first by Sasaki \cite{Sasaki:1987},
in synchronous gauge, and then by
Pyne and Birkinshaw \cite{Pyne:2003bn} in the Poisson gauge.  Bonvin, Durrer and Gasparini \cite{Bonvin:2005ps} carefully included all relativistic effects for scalar perturbations at first order, and this was extended to include vector and tensor modes in~\cite{DiDio:2012bu}.

Partial results towards the full second-order computation  of cosmological distance were given by Barausse, Matarrese and Riotto~\cite{Barausse:2005nf}, in the case of a matter-dominated universe. The full second-order result, for a universe with dark energy, was presented independently in our paper I \cite{Umeh:2012pn} and by Ben-Dayan et al~\cite{BenDayan:2012wi} (see also related work~\cite{BenDayan:2012ct,
BenDayan:2013gc,Fanizza:2013doa,Nugier:2013tca}).

Our result, when specialized to the partial formulas presented in \cite{Barausse:2005nf} and transformed to the comoving-synchronous gauge used in \cite{Barausse:2005nf}, is in agreement with \cite{Clarkson:2011uk}. The task of establishing agreement between our version of the full results and the version of \cite{BenDayan:2012wi} is highly nontrivial, because their work was done not only in another gauge, but also in a different coordinate system (geodesic lightcone coordinates). 
Here we give the detailed derivation of the results presented in paper I.

We assume a flat FLRW background; the extension to curved backgrounds is nontrivial and is left for future work. Vector and tensor modes at first order are neglected. However, we do incorporate second-order vector and tensor modes that are sourced by the product of first-order scalar fluctuations ~\cite{Lu:2007cj,Ananda:2006af,Lu:2008ju,Baumann:2007zm}. For the vector modes, this is particularly important because the amplitude of the vector potential can be of order 1\% of the first-order scalar potential. Tensor modes are included at second order for completeness. We allow for dynamical dark energy, with only one restriction, i.e. that we neglect anisotropic stress at first order. This excludes only a small subset of dark energy models. (Note that at first order after decoupling, anisotropic stress in CDM and baryons is negligible, while for photons and neutrinos the anisotropic stress makes a negligible contribution.) Anisotropic stress is generated at second order, and we include this. 

In general relativity, the field equations at first order imply equality of the metric perturbations in Poisson gauge, $\Psi=\Phi$, when the anisotropic stress vanishes. We adopt $\Psi=\Phi$ as an assumption at first order -- but we do not make any further use of the field equations at first or second order. Our perturbative expansions assume that the metric perturbations and peculiar velocity of matter are small, but we do not require the density contrast to be small, so that our results apply into the mildly nonlinear regime.

The assumption that $\Psi=\Phi$ at first order means that our results do not apply to general modified gravity theories that are an alternative to dark energy. When $\Psi\neq\Phi$ at first order, there are significant complications, and we leave this more general case for future work.

There are a variety of ways to calculate our result. We work mainly in a perturbed Minkowski spacetime, conformally transforming our key results at the end to FLRW spacetime. The key steps in our derivation are as follows (perturbatively evaluated order by order):
\begin{enumerate}
\item Solve the null geodesic equation for the photon 4-momentum as a function of background affine parameter. 
\item Calculate the redshift  as a function of affine parameter.
\item Solve the Sachs equations for the area distance as a function of affine parameter.
\item[$\rightarrow$] All quantities are now in terms of the affine parameter of the background. The distance-redshift relation at this stage is in the form of two parametric equations with the background affine parameter linking them.
\item Perturbatively invert the redshift-affine parameter relation and substitute into the area-distance relation.
\end{enumerate}
We then find an explicit expression for the area distance to a source at observed redshift $z_s$ as $D_A(z_s)$, given by~\eqref{bigboy}, which is written as a function of the background comoving distance $\chi_s$ to the source -- which is in turn determined using the background distance-redshift relation for the \emph{observed} redshift (more on this confusing issue in Section~\ref{sec:obs}). We write our result in terms of the metric potentials and peculiar velocity, all defined in the Poisson gauge. 

This paper is organised as follows: we provide general  covariant evolution equations for null shear, null expansion and the area distance in Section \ref{sec:Covframe-work}. The photon geodesic equations are derived and solved perturbatively in Section \ref{sec:geodesic-pert},  and we calculate the physical redshift in terms of the perturbed metric variables in Section \ref{sec:physical-red}. The area distance in perturbation theory is derived and solved order by order in Section \ref{sec:area-ditance}. We present a general nonlinear expression for expressing the area distance in terms of the physical redshift of the source in Section \ref{sec:obs}, which contains our main result (an alternative form is to be found in  Appendix~\ref{alternative}). 

{\em Notation:} We use indices $a,b,c,\cdots=0,1,2,3$ in a general spacetime. In perturbed FLRW, the indices $i,j,\cdots=1,2,3$ denote spatial components, and the linearly perturbed metric in Poisson gauge (scalar modes) is $a^2[-(1+2\Phi)\d\eta^2+(1-2\Phi)\delta_{ij}\d x^i\d x^j]$. The conformal Hubble rate is $\mathcal{H}=a'/a$.

\section{Preliminaries}

\subsection{Nonlinear description of distances and redshift}\label{sec:Covframe-work}

Here we present the key equations for light propagation in a general spacetime, which we will solve perturbatively . 
First we consider a light ray with tangent vector $k^a$ and affine parameter $\lambda$, on the past light cone, which is  a  constant phase hypersurface, ${S}=\,$const :
 \be
{ k}^{a} = \frac{\d { x}^{a}}{\d\lambda} \,,~~~{k}_a= {\nabla}_a  S.
 \ee
The tangent vector is null and geodesic:
\begin{equation}\label{Geodesiceqn1}
{k}_a{k}^a=0,\,\,\,\,\,  {k}^b{\nabla}_b{ k}^a=0\,,
\end{equation}
and may be decomposed relative to an observer with  4-velocity $u^a$ into parallel and orthogonal components:  
\begin{eqnarray}\label{4vector}
{ k}^a=(-{ u}_b{ k}^b)\left({ u}^a-{ n}^a\right)={E}\left({ u}^a-{ n}^a\right),~~~n_an^a=1,~n_au^a=0, ~{E}=-{ u}_b{ k}^b.
\end{eqnarray}
Here $n^a$ is the unit direction vector of observation, and $k^a$ is along an \emph{incoming} light ray on the past light cone of the observer. $E$ is the photon energy measured by $u^a$.  Note that our choice of $n^a$ is {\em opposite} to the direction of photon propagation.  

The screen space is orthogonal to the light ray and to the observer 4-velocity, and the tensor 
\begin{equation}\label{Screenspacemtric}
{N}_{ab}= {g}_{ab}+{u}_a{u}_b-{n}_a{n}_b\,,
\end{equation}
projects into screen space. It
satisfies the following relations
\begin{equation}
 {N}^a{}_a=2,
~~~ {N}_{ac}{{N}^c}_b={N}_{ab},
~~~{N}_{ab}{k}^b={N}_{ab}{u}^b={N}_{ab}{n}^b=0\,.
\end{equation}
For any  spatial tensor $T^{a\cdots }{}_{\cdots b}$, we can isolate the parts lying in the screen space and parallel to $n^a$: 
\begin{eqnarray}
{{T}_\o}^{a\cdots }{}_{\cdots b}&=&{N}^a{}_c\cdots {N}^d{}_b {T}^{c\cdots }{}_{\cdots d}\,,\\
{T}\p&=&{n}_a\cdots n^b\,{T}^{a\cdots }{}_{\cdots b}\,.
\end{eqnarray}
%

The invariant decomposition of the covariant derivative of the photon ray vector is given by,
\begin{equation}
 {\nabla}_b k_a= \frac{1}{2}\theta N_{ab}+ {\Sigma}_{ab}\,,
\end{equation}
where 
\begin{eqnarray}
{\theta}\equiv N^{ab} {\nabla}_a k_b,~~~~{\Sigma}_{ab}={\Sigma}_{\<ab\>}\equiv N_{(a}{}^c N_{b)}{}^d {\nabla}_c  k_d-\frac{1}{2}{\theta}N_{ab}\,.
\end{eqnarray}
Thus  ${\theta}$ describes the rate of expansion of the area of a bundle of light rays and ${\Sigma}_{ab}$ describes its rate of shear (the trace-free part of the derivative projected into the screen space). The angled brackets on indices denote the trace-free part of a screen space projected tensor. Note that there is no null vorticity since $k_a={\nabla}_aS$.

The Sachs propagation equations for the null shear and null expansion are~\cite{Clarkson:2011br},
\begin{eqnarray}\label{Raychau}
\frac{\d {\theta}}{\d \lambda}&=&-\frac{1}{2}{\theta}^2-  {\Sigma}_{ab}\Sigma^{ab}- R_{ab} k^a k^b\,, \\
\label{eq:shearevo}
\frac{\d{\Sigma}_{ab}}{\d\lambda}&=&-{\Sigma}_{ab}{\theta}+ C_{acbd}  k^c k^d\,,
\end{eqnarray}
where ${\text{D} }/{\d\lambda}= k^a{ \nabla}_a $. At this point one may use the Einstein equations to replace the Ricci tensor in  (\ref{Raychau}), but we keep these expressions general in this work, and write them in terms of the metric. The invariant area of bundle in screen space, $\cal A$,  defines the angular diameter (or area) distance $ D_A$, and is directly related to the null expansion: 
\begin{equation}\label{AreaExpansion}
\frac{1}{\sqrt{\cal A}}\frac{\d\sqrt{\cal A}}{\d\lambda}=\frac{\d \ln {D}_A}{\d \lambda}=\frac{1}{2}{\theta}\,.
\end{equation}
Substituting   (\ref{AreaExpansion}) in   (\ref{Raychau}), we obtain a second order differential equation for the area distance, 
\begin{equation}\label{CovAreadistance}
\frac{\d^2 D_A}{\d\lambda^2}=-\frac{1}{2} \left[ R_{ab} k^a k^b+{\Sigma_{ab}}\Sigma^{ab}\right] {D}_A.
	\end{equation}

The affine parameter associated with $k^a$ is not an observable, but it is related to the redshift of the photon via \eqref{Redshiftprorpa1} below. The redshift is given by
\begin{equation}
\label{eq:redshift}
1+ z = {(- k_a u^a)_{s} \over (- k_b u^b)_{o}}={ E_s\over  E_o} \,, 
\end{equation}
where  `$s$' denotes the source (e.g., a galaxy) and `$o$' the observer.
The covariant derivative of the 4-velocity is invariantly decomposed as:
 \begin{eqnarray}
 {\nabla}_{b} u_{a}=- A_{a}u_{b}+\frac{1}{3} \Theta ( g_{ab }+u_au_b) 
+ \sigma_{ab}+  \Omega_{ab}\,,
 \label{du}
\end{eqnarray}
where $ \Theta$ is  the volume expansion rate of the $u^a$ worldlines,   $ A_a$ is the 4-acceleration,   $ \sigma_{ab}$ is the shear tensor and $\Omega_{ab}$ is the vorticity tensor. 
In terms of these variables, the photon geodesic equation~(\ref{Geodesiceqn1}) reduces to the equations for the photon  energy and  observational direction  
\begin{eqnarray}\label{energyprorpa1}
\frac{\d  E}{\d \lambda}&=&- E^2\left[\frac{1}{3}\Theta -  A_{a}n^{a}+ \sigma_{ab} n^{a} n^{b}\right]\,,
\\
\frac{\text{D}n^a }{\d\lambda}&=&E \left[n^a\left(A_bn^b-\sigma_{bc}n^bn^c \right) -A^a+ \left(\sigma^a{}_b-\Omega^a{}_b\right) n^b
+ u^a\left(\frac{1}{3}\Theta - A_bn^b + \sigma_{bc}n^bn^c\right)\right].
\end{eqnarray}
Using (\ref{energyprorpa1}), the  redshift propagation equation becomes,
\begin{eqnarray}\label{Redshiftprorpa1}
\frac{\d  z}{\d \lambda}=-(1+ z)^2\left[\frac{1}{3} \Theta -  A_{a}n^{a}
+ \sigma_{ab}n^{a}n^{b}\right]\,.
\end{eqnarray}
This is a general non-perturbative and coordinate-independent propagation equation for the  observed redshift. 

Finally, we require appropriate boundary conditions for \eqref{CovAreadistance}. These can be found from the a series expansion for distances [Kristian and Sachs \cite{KandS66}, equation (34)]:
\be
D_A^2(\lambda_s)=(u_a k^a)^2_o(\lambda_o-\lambda_s)^2\left[1-\frac{1}{6}(R_{ab}k^ak^b)_o(\lambda_o-\lambda_s)^2+\cdots\right],
\ee
which implies
\be\label{conf2}
D_A(\lambda_o)=0~~~\mbox{and}~~~\frac{\d D_A}{\d\lambda}\bigg|_o=-E_o .
\ee
In the subsequent sections, we compute  in cosmological perturbation theory the equtions (\ref{eq:shearevo}), (\ref{CovAreadistance}) and \eqref{Redshiftprorpa1}, with boundary conditions at the origin from \eqref{conf2}.

\subsection{Cosmological perturbation theory at second order}

\begin{figure}
\begin{center}
\includegraphics[width=0.69\textwidth]{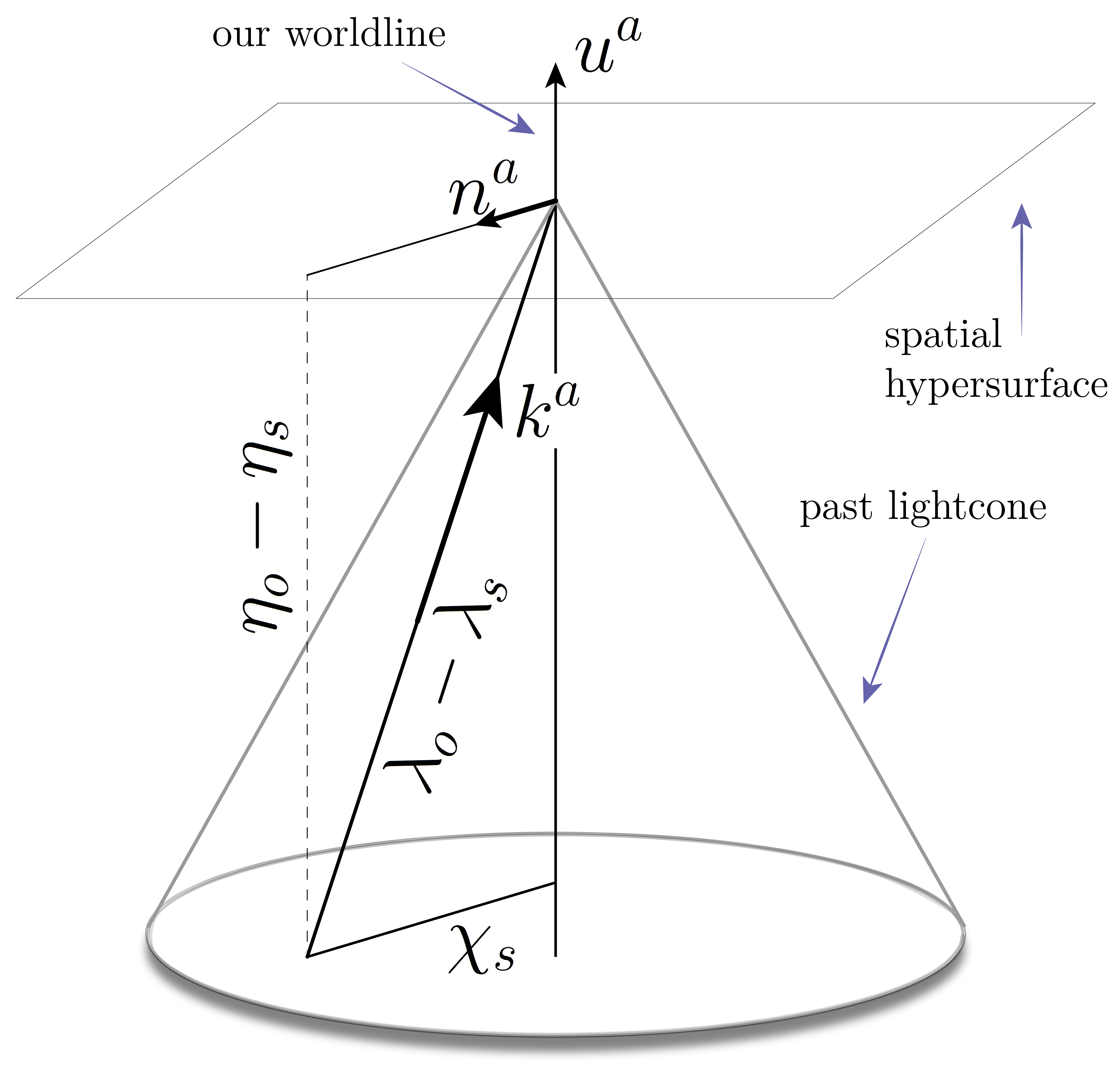}
\caption{Spacetime diagram showing the observational configuration.}
\label{spacetime}
\end{center}
\end{figure}

From now on we consider perturbations about a flat FLRW metric to second-order, using two metrics: the physical spacetime metric $\hat g_{ab}$, and a conformally related one, $g_{ab}=a^{-2}\hat g_{ab}$, where $a$ is the scale factor of the background. We use a hat to denote quantities living on the physical spacetime so that quantities on the conformal spacetime  have no hat. 
The background of the metric $g_{ab}$ is Minkowski, which offers a major simplification of the equations.  

Conformal maps preserve both angles and shapes of infinitesimally small figures, but not their size. 
The area distance transforms as $\hat{D}_A=a D_A$.  The conformal transformation  $\hat{g}_{ab}\rightarrow g_{ab}$  maps the null geodesic equation of the perturbed FLRW metric $\hat g_{ab}$ to a null geodesic on the perturbed Minkowski metric $ g_{ab}$. The affine parameters  transform as $\d \hat \lambda \rightarrow \d \lambda = a^{-2} \d \hat{\lambda}$, so that the photon ray vector transforms as $\hat k^b=a^{-2}k^b\Leftrightarrow\hat k_a=k_a$. For a 4-velocity, we have $\hat u^a=a^{-1}u^a\Leftrightarrow \hat u_a=au_a$ .  Hence, the photon energy transforms as $\hat E= -\hat u_b\hat k^b = -a^{-1} \,u_b k^b=a^{-1} E$.  (We normalize $E=1$ in the Minkowski background.) In summary:
\bea
&&\mbox{physical (perturbed FLRW) metric}~\hat{g}_{ab}~\leftrightarrow~\mbox{conformal (perturbed Minkowski) metric}~g_{ab}=a^{-2}\hat g_{ab},\nonumber\\
&& \hat{D}_A=a D_A,~~\hat\lambda=a^2\lambda,~~\hat{k}_a=k_a,~~\hat {u}^a=a^{-1}u^a,~~\hat E= a^{-1}E. \label{conf3}
\eea
A general tensor in the conformal spacetime $g_{ab}$ is expanded as
\be
 \tot{T}=\back{T}+\delta T+\frac{1}{2}\delta^2T~~\mbox{or}~~ T= T^{(1)}+{1\over2}T^{(2)}\,,
\ee
where the second relation applies to
quantities that vanish in the background. For convenience, we will simplify the notation by omitting (a)~the overbar on the background quantity when this is clear from the context; (b)~the $(1)$ in the first-order part $T^{(1)}$ (e.g. $\Phi$ as short-hand for $\Phi^{(1)}$), when this does not lead to confusion; (c)~the $(2)$ superscript on vector and tensor quantities that vanish at zero and first order.
We use the same convention for perturbations of objects in the `full' (expanding) spacetime~-- these only differ by factors of $a$ for the most part.

For a flat background the physical metric in the Poisson gauge is
\begin{eqnarray}\label{metric}
\d \hat s^2 &=&
a^2\d s^2 = a^2\big[-\big\{1 + 2\Phi+\Phi\two \big\}\d \eta^2 + 2\omega_{i} \d x^{i}\d \eta
 + \big(\big\{1-2 \Phi-\Psi\two\big\}\gamma_{i j} + h_{ij}\big)\d x^{i}\d x^{j}\big]\,, \\ \nonumber
\gamma_{ij} &=& \delta_{ij}~~\mbox{if $x^i$ are cartesian.} 
\end{eqnarray}
Here $\gamma_{ij}$ is the spatial metric of the Minkowski background in general spatial coordinates,  and  $\nabla_i$ is its covariant derivative ($\nabla_i=\partial_i$ in cartesian coordinates). 

In the Minkowski background, we define derivatives along and transverse to the radial direction: 
\bea
\D\p=n^i\D_iX_{j\cdots},~~~~~~\nabla_{\o i}X_{j\cdots}=(\gamma_{i}^{~k}-n_in^{k} )(\gamma_{j}{}^l-n_jn^{l} )\cdots\nabla_{k} X_{l\cdots}\,,
\eea
where $n_i=\nabla_i\chi$. 
Then the radial derivative can be interchanged with the derivative along a null geodesic:
\begin{equation}\label{eq:therelation}
\D\p X=X'-\frac{\d X}{\d\lambda}\equiv (\partial_\eta - d_\lambda) X \,,
\end{equation}
which we consistently use to remove radial derivatives. 
On the background one can replace the affine parameter $\lambda$ with the conformal time $\eta$, and both are also related to the radial distance along the past lightcone: $\lambda_o-\lambda=\eta_o-\eta= \chi$. Useful identites involving derivatives and integrals are given in Appendices A and B.

In \eqref{metric}, $\Phi$ is the first-order scalar potential, and $\omega_i$ and $h_{ij}$ are the second-order vector and tensor contributions. By setting $\Phi=\Psi$ at first-order, we neglect first-order anisotropic stress. We also neglect first-order vector and tensor modes. Other than this, our results are general, and do not assume any form of matter. For convenience the standard results for the second-order potentials for a LCDM model are given in Appendix C, but we do not use these here.  

The physical 4-velocity $\hat u^a$ is
\bea
\hat u^0 &=&\frac{1}{a}\left[ 1 - \Phi -\frac{1}{2}{}\Phi\two + \frac{3}{2}\Phi^2 + \frac{1}{2}\nabla_iv\nabla^iv \right],
\label{o2u0}\\
\hat u^i&=&\frac{1}{a}\left[ \nabla^iv+\frac{1}{2}\bigg\{\nabla^iv\two+v\twod^i\bigg\}\right],\label{o2ua}
\eea
where $v$ is the first-order scalar velocity potential, $v\two$ is its second-order part and $v\two_i$ is the second-order vector mode of the velocity.

\section{Null geodesics and the redshift}\label{sec:geodesic-pert}

Here we present solutions to the geodesic equations up to second order for the metric~\eqref{metric}. In general
\be
\tot{k}^a=\back{k}^a+\delta k^a+\frac{1}{2}\delta^2k^a\,,
\ee
and similarly for the physical spacetime.
We do not consider perturbations of the affine parameter explicitly, so that $\lambda$ is always a background quantity. In a perturbed equation at a given order indices are explicitly raised and lowered with the background metric. On the Minkowski background  we take $\bar\nabla_a \bar k_b=0$.

Perturbing the geodesic equation on the Minkowski background  up to second order gives (using $\text{D}/\d\lambda=\bar k^a\nabla_a$, with $\nabla_a$ always the background covariant derivative),
  \begin{eqnarray}
&&\mathcal{O}(0)~~~~\frac{\text{D} k^a}{\d\lambda}=0\,,\\
   \label{eq:geodesic1}
&&\mathcal{O}(1)~~~~\frac{\text{D}\delta k^{a}}{d\lambda}=-\delta k^a \nabla_{b} k^b-k^bk^c\delta\Gamma_{cd}^{a}= -\delta k^a \nabla_{b} k^b+  \frac{1}{2} k^{b} k^{c} {\nabla}^{a}\delta g_{bc} - k^{b} k^{c} \bar{g}^{ad}{\nabla}_ {c}\delta g_{bd}\,,\\
\label{eq:geodesic2}
&&\mathcal{O}(2)~~~~ \frac{\text{D}\delta^2{k}^{a}}{{d\lambda}}= -\delta^2 k^a \nabla_{b} k^b-k^b k^c\delta^2\Gamma^a_{bc}-4k^b \delta k^c \delta\Gamma_{bc}^a-2\delta k^b\nabla_b \delta k^a\,, \nonumber\\&&
~~~~~~~~~~~~~~~~~~~= -\delta^2 k^a \nabla_{b} k^b+ \frac{1}{2} k^{b} k^{c} {\nabla}^{a}\delta^2 g_{b c}- k^{b} k^{c}\bar{g}^{ad} {\nabla}_{c}\delta^2 g_{bd}
 - 2\delta k^{b} k^{c} \left[\bar{g}^{ad}{\nabla}_{b}\delta g_{cd}  
 - {\nabla}^{a}\delta g _{b c} +  \bar{g}^{ad}{\nabla}_ {c}\delta g_{bd} \right] \nonumber \\ &&
 ~~~~~~~~~~~~~~~~~~~-  \bar{g}^{ad}\bar{g}^{ef}\delta g_{df}
k^{b} k^{c} \left[{\nabla}_{d}\delta g_{b c} -  2 {\nabla}_{c}\delta g_{b d}\right] 
  -2 \delta k^{b} {\nabla}_ {b}\delta k^{a},
\end{eqnarray}
where $\nabla_{a}\bar{g}_{bc}=0$. We have substituted the perturbed Christoffel symbols with the corresponding perturbed metric and then the partial derivatives of the perturbed metric with the covariant derivatives of the background spacetime. (All terms containing the background Christoffel symbols cancel out -- we have not assumed Cartesians.) With the geodesic equation in this form, it is then straightforward to substitute the corresponding metric variables from~\eqref{metric}.

\subsection{First-order geodesic equation}

Substituting  \eqref{metric} in  \eqref{eq:geodesic1} gives 
\begin{eqnarray}\label{eq:geodesicsub1}
\frac{\text{d}\delta k^a}{\d \lambda}
&=&-2\left[u^a\left(\d_\lambda\Phi-\Phi'\right)+n^a\Phi'+\nabla_{\o}^a\Phi\right],
\end{eqnarray}
where we decomposed into parts along $u^a$, along $n^a$ and in the screen space. We have also substituted for $\D\p$ in terms of $\partial_\eta$ and $\d_\lambda$ using \eqref{eq:therelation}, a practice we follow below.
This may be integrated along the line of sight from observer to source. The time, radial, and screen-space parts are  
\begin{eqnarray}\label{eq:solngeodesicalong}
\delta k^0&=&
-2\Phi\big|_o^s + 2\I\Phi'\,,\\
\label{eq:solngeodesicp}
\delta k\p&=&-2\I\Phi'\,,\\
\delta k_{\o}^i&=& -2\I \nabla_{\o}^i \Phi\,.
\label{eq:solngeodesicpep}
\end{eqnarray}
Here and below we are using notation $\delta k\p=n^i\delta k_i$ and $\delta k_{\o}^i=N^{ij}\delta k_j$.  (As written~\eqref{eq:solngeodesicpep} is valid only in Cartesian coordinates because we have integrated over $\text{D}/d\lambda$ not $d/d\lambda$~-- this can be converted to arbitrary spatial coordinates if the $i$ index is reinterpreted as a tetrad index in the screen space. This restriction drops out when we calculate the redshift below~-- a scalar.) 
As with future equations of this form, the quantities are understood to be functions of $\lambda_s$, where $\lambda_s$ is calculated at the background position of the source. (More on this see appendix \ref{Notation}.) Note that $\delta k^a_o=0$.

The perturbative correction to the observation direction at first order is
   \begin{eqnarray}
    \delta n^i =-n^i\Phi\big|^s_o + 2 \I\nabla_\o^i\Phi\,.
    \label{delta_ni}
    \end{eqnarray}
The second term is effectively an integral over the transverse velocities along the line of sight. 

\subsection{Second order geodesic equation}

At second order we have four different kinds of terms contributing: induced scalars, induced vectors, induced tensors and terms explicitly quadratic in first order terms.
We use `induced' to describe scalars, vectors and tensors at second order because they are all sourced by first-order scalars squared in our context, i.e. where vectors and tensors at first order are set to zero.  For more details on Scalar-Vector-Tensor (SVT) decomposition see appendix \ref{Notation2}.

\begin{description}
\item[\tt{Induced scalars}] The geodesic equation for scalars at second order is exactly the same as in the first order case, but now we have the distinction between the potentials: 
\begin{eqnarray}\label{eq:geodesicsub2}
\frac{\text{d}\delta^2\Sd k^a}{\d \lambda}= u^a\left[\Phi'\twod+\Psi'\twod -2\d_{\lambda}\Phi\two\right] +n^a \left[\d_{\lambda}\big\{\Phi\two-\Psi\two\big\}-\big\{\Phi'\twod+\Psi'\twod\big\}\right]-\nabla^a_{\o}\big[ \Phi\two +  \Psi\two\big].
\end{eqnarray}
Integrating we have:
\begin{eqnarray}\label{eq:solngeodesictime}
\delta^2\Sd k^0&=&-2\Phi\two\big|_o^s +\I(\Phi'\twod+\Psi'\twod)\,,\\
\delta^2\Sd k\p&=& \Phi\two\big|_o^s-\Psi\two\big|_o^s - \I\left(\Phi'\twod+\Psi'\twod\right)\,,\\
\label{eq:solngeodesicradial}
\delta^2\Sd k_{\o }^i &=& - \I \nabla^i_{\o } \left(\Phi\two+ \Psi\two\right)\,.
\label{eq:solngeodesicspace2}
\end{eqnarray}

\item [\tt{Induced vectors}]  The geodesic equation for the induced vector contribution is  
\begin{eqnarray}\label{eq:geodesicvector}
\frac{\text{d} \delta^2\Vd k^a}{\d\lambda}&=& u^a\left(\omega'\p-\d_{\lambda}\omega\p\right)-n^a\omega'\p  -\left({\d_\lambda\omega^a_\o} + \nabla^a_\o \omega\p\right),
\end{eqnarray}
with solutions
\begin{eqnarray}\label{eq:solngeodesicvectortime}
\delta^2\Vd k^0&=& 
-\omega\p\big|_o^s+ \I{\omega'\p}\,,\\
\delta^2\Vd k\p &=&-\I\omega'\p \,,\\
\label{eq:solngeodesicvectorradial}
\delta^2\Vd k_{\o}^i&=& -{\omega_\o}^i\big|_o^s- \I\nabla_{\bot}^i\omega\p\,.
\label{eq:solngeodesicvectorspatial}
\end{eqnarray}

\item [\tt{Induced tensors}] 
The geodesic equation for the induced tensor perturbations is  
\begin{eqnarray}
\frac{\text{d}\delta^2\Td k^a}{\d \lambda}&=&-h'\p u^a+ n^a\left(\d_{\lambda}h\p+h\p'\right)+2\d_{\lambda}h^a_\po+\nabla^a_\o h\p,
\end{eqnarray}
with solutions
\begin{eqnarray}
\delta^2\Td k^0&=&
 -\I h'\p\,,\\
\delta^2\Td k\p &=& h\p\big|_o^s + \I h'\p \,,\\
\delta^2\Td k_{\o}^i&=& 2  h_{\po}^i\big|_o^s + \I\nabla_{\bot}^{i}h\p \,.
\end{eqnarray}

\item [\tt{Nonlinear quadratic terms}]
The geodesic equation for terms quadratic in the first-order gravitational potential is 
\begin{eqnarray}\label{eq:NLgeoodesic}
\frac{\text{d}\delta^2\NLd k^a}{\d\lambda}&=&u^a\left[-2\left(\delta k^0+\delta k_{\pp}\right)\left(\D_{\pp}\delta k^0+2\d_{\lambda}\Phi\right)+ 4\left(2\delta k_{\pp}-\delta k^0+2\Phi\right)\left(\d_{\lambda}\Phi-\Phi'\right)-2\delta k_{\o}^i\left(\nabla_{\o i}\delta k^0-2\nabla_{\o i}\Phi\right)\right]\nonumber\\ \nonumber&&
+n^a\left[-2\left(\delta k^0+\delta k_{\pp}\right)\left(\D_{\pp}\delta k_{\pp}-2\d_{\lambda}\Phi\right)-4\Phi'\left(\delta k^0+2\Phi\right)- 2\delta k^i_{\o}\left(\nabla_{\o i}\delta k_{\pp}+2\nabla_{\o}\Phi\right)\right]\\ &&
+\left[-2\D_{\pp}\delta k^a_{\o}\left(\delta k^0+\delta k_{\pp}\right)-2\delta k_{\o}^a\left[(\lambda_o-\lambda)^{-1}\delta k_{\pp}-2\d_{\lambda}\Phi\right]-2\delta k^i_{\o}\nabla_{\o i}\delta k^a_{\o}+4\nabla_{\o}^b\Phi\left(\delta k_{\pp}-2\Phi\right)\right]\,.
\end{eqnarray}
Substituting from \eqref{eq:solngeodesicalong}, \eqref{eq:solngeodesicp} and \eqref{eq:solngeodesicpep} gives
\begin{eqnarray}\label{eq:Qgeodesiceqtime}
\frac{1}{8}\frac{\d\delta^2\Q k^0}{\d\lambda}&=&\left(-\Phi_o+2\Phi-\I\Phi'\right)\left({\d_{\lambda}}
\Phi-\Phi'\Phi\right)
+\I\frac{(\lambda_o-\lambda)}{(\lambda_o-\lambda_s)}\nabla_{\o}^i\Phi'\I\nabla_{\o i}\Phi\,,\\
\frac{1}{8}\frac{\d\delta^2\Q k_{\pp}}{\d\lambda}&=&-2\Phi'\Phi_o-\d\lambda \Phi (\Phi-\Phi_o)+\Phi'\left(\Phi-\I\Phi'\right)+\left[\nabla_{\o}^i\Phi
-\I\frac{(\lambda_o-\lambda)}{(\lambda_o-\lambda_s)}\nabla_{\o}^i\Phi'\right]\I\nabla_{\o i}\Phi\,,\\
\label{eq:Qgeodesiceqradial}
\frac{1}{8}\frac{\text{d}\delta^2\Q k^i_{\o}}{\d\lambda}&=&\Phi_o\nabla_{\o}^i\Phi-\nabla_{\o}^i\Phi\I\Phi'
-\left((\lambda_o-\lambda)^{-1}\I\Phi'+\d_\lambda\Phi\right)\I\nabla_{\o}^i\Phi-\I\nabla_{\o}^k\Phi\I\frac{(\lambda_o-\lambda)}{(\lambda_o-\lambda_s)}\nabla_{\o k}\nabla_{\o}^i\Phi\,.
\label{eq:Qgeodesiceqscreenspace}
\end{eqnarray}
On integrating, the quadratic terms can be separated into their SW (boundary), ISW and integrated ISW (IISW) parts:
 \begin{eqnarray}
  \frac{1}{8} \delta^2\Q k^0&=&\Phi_s\Phi|_{o}^s-\left(\Phi_s-\Phi_o\right)\I\Phi'-\I\Phi\Phi'
+\I\Phi'\Ipr\Phi'(\tilde{\lambda})
+\I
\Ipr\frac{(\lambda_o-\tilde{\lambda})}{(\lambda_o-\lambda)}\nabla_{\o}^i\Phi'(\tilde{\lambda})\Ipr\nabla_{\o i}\Phi(\tilde{\lambda})\,,\label{dqk0}\\
\frac{1}{8}\delta^2\Q k_{\pp}&=&-\frac{1}{2}\left(\Phi_s^2-\Phi_o^2\right)-2\Phi_o\I\Phi'+\I\Phi\Phi'
-\I\Phi'\Ipr\Phi'(\tilde{\lambda})
+\I \nabla_{\o}^i\Phi\Ipr\nabla_{\o i}\Phi(\tilde{\lambda})
 \nonumber \\ &&
-\I\Ipr\frac{(\lambda_o-\tilde{\lambda})}{(\lambda_o-\lambda)}\nabla_{\o}^i\Phi'(\tilde{\lambda})\Ipr\nabla_{\o i}\Phi(\tilde{\lambda})\,,\\
\frac{1}{8}\delta^2\Q k^i_{\o}&=&-(\Phi_s+\Phi_o)\I\nabla_{\o}^i\Phi+\I\Phi\nabla_{\o}^i\Phi
-\I\nabla_{\o}^i\Phi\Ipr\Phi'(\tilde{\lambda})
-\I(\lambda_o-\lambda)^{-1}\I\Phi'(\tilde{\lambda})\I\nabla_{\o}^i\Phi(\tilde{\lambda})
\nonumber \\ &&
-\I\Ipr\nabla_{\o}^k\Phi(\tilde{\lambda})\Ipr\frac{(\lambda_o-\tilde{\lambda})}{(\lambda_o-\lambda)}\nabla_{\o k}\nabla_{\o}^i\Phi(\tilde{\lambda})\,.
   \end{eqnarray}
This is useful for identifying the physical meaning of each term. We therefore have the full second-order contribution to the ray vector $k^a$ as
\be\label{o2ka}
\delta^2k^a=\delta^2\Sd k^a+\delta^2\Vd k^a+\delta^2\Td k^a+\delta^2\NLd k^a\,.
\ee

\end{description}

\subsection{Observed redshift}
\label{sec:physical-red}

We now expand the photon energy, $E=-u^ak_a$,  up to second order,
$\tot{E}=\back{E}+ \delta E + \delta^2 E/2$.
Following \eqref{o2ka}, we decompose $\delta^2 E$ as
\begin{eqnarray}
\delta^2 E= \delta^2\Sd E+\delta^2\Vd E+ \delta^2\Td E + \delta^2\NLd E\,.
\end{eqnarray}
Using \eqref{o2u0}, \eqref{o2ua}, \eqref{eq:solngeodesicalong}--\eqref{eq:solngeodesicpep} and \eqref{o2ka}, we find
  \begin{eqnarray}
 \delta E &=& \delta k^0+  \Phi + \D\p  v \,,\\
   \delta^2\Sd E&=&\delta^2\Sd k^0+\Phi\two+   \D\p  v\two \label{o1e}
   \,,\\
  \delta^2\Vd E&=&\delta^2\Vd k^0+ \omega\p \two +
  v\p \two
  \,,\\
   \delta^2\Td E&=&\delta^2\Td k^0\,,\\
     \delta^2\NLd E 
    &=&\delta^2\NLd k^0 +2\Phi \delta k^0 -2\delta k{\p}\D{\p} v-2\delta k_{\o}^i\nabla_{\o i}v-\Phi^2 -4 \Phi\D{\p} v+\D{\p}v \D{\p} v+\nabla_{\o i}v\nabla_{\o}^iv\,.
\end{eqnarray}
The perturbed photon vector at the observer is zero, so that  
\begin{eqnarray}
 \delta E_o &=&   \Phi_o + \D\p  v_o \,,\\
   \delta^2\Sd E_o&=&\Phi_o\two
   + \D\p  v_o\two
  \label{eq:deltasqEobsS} \,,\\
  \delta^2\Vd E_o&=& {\omega}_{\pp o}\two +
  v_{\pp o}\two
 \label{eq:deltasqEobsV} \,,\\
   \delta^2\Td E_o&=&0\,,\\
     \delta^2\NLd E_o &=&
 -\Phi^2_o
  -4 \Phi_o  \D\p v_o
  +\D_iv_o\D^iv_o\,.
 \label{eq:deltasqEobsQ}
\end{eqnarray}

The  observed redshift in  (\ref{eq:redshift}) may also be expanded: 
\begin{eqnarray}\label{eq:reshiftexp}
a(1+\hat{z})=1+\delta z +\frac{1}{2}\delta^2 z =\frac{{E}_s}{{E}_o}&=& \frac{k^au_a|_s}{k^bu_b|_o}
=\frac{\left(\back{E}_s+ \delta E_s +\frac{1}{2} \delta^2 E_s\right)}{\left(\back{E}_o+\delta E_o + \frac{1}{2}\delta^2 E_o\right)}.
\end{eqnarray}
Note we define perturbed redshift $\delta z$ relative to the Minkowski background, where the background value is zero.
Using $\bar E_s=\bar E_o=1$, this becomes
\begin{eqnarray}\label{eq:reshiftexp2}
a({1+\hat{z}})&=&1+ \left({\delta E_s} -{\delta E_o }\right)
+ \left[ \frac{1}{2}{\delta^2 E_s}-\frac{1}{2}{\delta^2 E_o} +{\delta E_o^2}
- {\delta E_o}{\delta E_s} \right]\,.
\end{eqnarray}
Now we may calculate the  redshift up to second order by substituting for $k^a$ in the equations above. 
The full contributions to the redshift of a source $s$ are as follows.
 
\begin{description}

\item [\tt{First order}] At first order, 
\begin{eqnarray}\label{eq:redshift1}
{\delta z}
&=&\left(\D\p v-\Phi\right)\big|^s_o +2\I  \Phi'\,.
\end{eqnarray}
This contains: the  Doppler term, which quantifies the effect of the velocity difference between the source and the observer; the SW term,  quantifying the difference in the gravitational potential between source and observer; and the usual ISW term.

\end{description}

\noindent At second order we have the following contributions: 
\begin{eqnarray}\label{red2nd}
\delta^2 z= \delta^2\Sd z+\delta^2\Vd z + \delta^2\Td z + \delta^2\NLd z\,.
\end{eqnarray}

\begin{description}

\item [\tt{Induced scalars}] Substituting \eqref{eq:solngeodesictime}--\eqref{eq:solngeodesicspace2}  into~\eqref{eq:reshiftexp2},
\begin{eqnarray}
{\delta^2\Sd z }=
\left(\D\p  v\two-\Phi\two\right)\big|^s_o
+\I\left( \Phi\twod' + \Psi\twod'\right) \,.
\end{eqnarray}

\item [\tt{Induced vectors}] We use  \eqref{eq:solngeodesicvectortime} in  \eqref{eq:reshiftexp2} to give,
\begin{eqnarray}\label{eq:reshifVec}
{\delta^2\Vd z}=\left(v\p\two+\omega\p\two  \right)\big|^s_o
+\I{\omega\p\two}'\d\lambda\,.
\end{eqnarray}

\item [\tt{Induced tensors}] 
\begin{eqnarray}
{\delta^2\Td z}=-\I {h\p\two}'  \,.
\end{eqnarray}

\item [\tt{Nonlinear quadratic terms}]
The contribution to the observed redshift from terms quadratic in the first-order gravitational potential is more complicated:
\begin{eqnarray}\label{eq:nonlinearred}
{\delta^2\NLd z}=\delta^2\NLd E_s-\delta^2\NLd E_o  +2\delta E_o^2
- 2\delta E_o \delta E_s \,.
\end{eqnarray}
We decompose $\delta^2\NLd E$ as 
\begin{eqnarray}\label{eq:ENLequation}
  \delta^2\NLd E &=&\delta^2\NLd k^0 
  - \Phi^2 - 2 \Phi\delta k^0 -4\Phi \D\p v-\D\p v\delta k\p - 2 \delta k_{\bot i}\nabla_{\bot}^i v
  + \D\p  v\D\p v+ \nabla_{\bot i}v\nabla_{\bot}^i v\,,
  \end{eqnarray}
where $\D_iv\D^iv=\D\p v\D\p v + \nabla_{\bot k}v\nabla_{\bot }^k v $ and $\delta^2\NLd k^0$ is given by  \eqref{dqk0}. For $\delta^2\NLd E_o$ we use \eqref{eq:deltasqEobsQ}. Then  \eqref{eq:ENLequation} becomes 
   \begin{eqnarray}
  \delta^2\NLd E_s &=&
  \Phi_s(3\Phi_s-4\Phi_o)-4\Phi_s\D\p v_s + \D\p v_s \D\p v_s
  +\nabla_{\bot k}v_s\nabla_{\bot }^k v_s + 4\left(2\Phi_o- \Phi_s+\D\p  v_s\right)\I \Phi' \nonumber \\ && 
  + 4 \nabla_{\bot i} v_s \I \nabla_{\bot}^i \Phi   
-8\I\Phi\Phi'+8\I\Phi'\Ipr\Phi'(\tilde{\lambda})+8\I
\Ipr\frac{(\lambda_o-\tilde{\lambda})}{(\lambda_o-\lambda)}\nabla_{\o}^i\Phi'(\tilde{\lambda})\Ipr\nabla_{\o i}\Phi(\tilde{\lambda})\,. \label{eq:ENLsource}
  \end{eqnarray}
At the observer,
\begin{eqnarray}\label{eq:ELobs2}
  \delta^2\NLd E_o&=&
 -\Phi^2_o
  -4 \Phi_o  \D_{\| }v_o
  +\D\p v_o \D\p v_o+\nabla_{\bot k}v_o\nabla_{\bot }^k v_o\,.
  \end{eqnarray}
Using  \eqref{o1e},   \eqref{eq:ENLsource} and \eqref{eq:ELobs2}  in \eqref{eq:nonlinearred}, we obtain the general form of quadratic terms contributing to the observed redshift:
  \begin{eqnarray}
  \delta^2\Q z_s&=&(\Phi_s-\Phi_o)\left(3\Phi_s+\Phi_o\right)-4\left(\Phi_s\D\p  v_s- \Phi_o  \D_{\| }v_o\right)+2 \left( \Phi_s\D\p  v_o- \Phi_o\D\p v_s\right)+ \left(\D\p v_o-\D\p v_s\right)^2
   \nonumber\\ &&
 +\nabla_{\bot k}v_s\nabla_{\bot }^k v_s-\nabla_{\bot k}v_o\nabla_{\bot }^k v_o
  +4\left[\left(\D\p v_s-\D\p v_o\right)-\left(\Phi_s-\Phi_o\right)\right] \I\Phi'+4 \nabla_{\bot i} v_s \I \nabla_{\bot}^i \Phi
   \nonumber\\ &&
  -8\I\Phi\Phi'
 +8\I\Phi'\Ipr\Phi'(\tilde{\lambda})+8\I
\Ipr\frac{(\lambda_o-\tilde{\lambda})}{(\lambda_o-\lambda)}\nabla_{\o}^i\Phi'(\tilde{\lambda})\Ipr\nabla_{\o i}\Phi(\tilde{\lambda})\,.
  \end{eqnarray}
 It is a sum of products of Doppler, SW and ISW terms both parallel and perpendicular to the line of sight:
\begin{eqnarray}
{\delta^2\NLd z }&=&\delta^2\NLd z_{\text{\tiny SW}}+ \delta^2\NLd z_{\text{\tiny SW}\times \text{\tiny ISW}}+ \delta^2\NLd z_{\text{\tiny Dop}\p }+ \delta^2\NLd z_{\text{\tiny Dop}_{\bot}}+\delta^2\NLd z_{\text{\tiny SW} \times \text{\tiny Dop}\p }+\delta^2\NLd z_{\text{\tiny Dop}\p \times \text{\tiny ISW}\p }+\delta^2\NLd z_{\text{\tiny Dop}_{\bot}\times \text{\tiny ISW}_{\bot}} 
+\delta^2\NLd z_{\text{\tiny IISW} },
\end{eqnarray}
where  
  \begin{eqnarray}
\delta^2\NLd z_{\text{\tiny SW}}&=&(\Phi_s-\Phi_o)\left(3\Phi_s+\Phi_o\right)\,,\\
\delta^2\NLd z_{\text{\tiny SW} \times \text{\tiny{ISW}}}&=&-4\left(\Phi_s-\Phi_o\right) \I\Phi'\,,\\
\delta^2\NLd z_{\text{\tiny SW} \times \text{\tiny Dop}\p }&=&-4\left(\Phi_s\D\p  v_s- \Phi_o  \D_{\| }v_o\right)
+2 \left( \Phi_s\D\p  v_o- \Phi_o\D\p v_s\right)\,,\\
\delta^2\NLd z_{\text{\tiny Dop}\p }&=&
\left(\D\p v_o-\D\p v_s\right)^2\,,\\
\delta^2\NLd z_{\text{\tiny Dop}_{\bot}}&=&\nabla_{\bot k}v_s\nabla_{\bot }^k v_s-\nabla_{\bot k}v_o\nabla_{\bot }^k v_o\,,\\
\delta^2\NLd z_{\text{\tiny Dop}\p \times \text{\tiny ISW}\p }&=&4\left(\D\p v_s-\D\p v_o\right) \I\Phi'\,, \\
\delta^2\NLd z_{\text{\tiny Dop}_{\bot}\times \text{\tiny ISW}_{\bot}} &=& 4 \nabla_{\bot i} v_s \I \nabla_{\bot}^i \Phi\,, \\
 \delta^2\NLd z_{\text{\tiny IISW}}&=& -8\I(\Phi\Phi')
 +8\I\Phi'\Ipr\Phi'(\tilde{\lambda})+8\I
\Ipr\frac{(\lambda_o-\tilde{\lambda})}{(\lambda_o-\lambda)}\nabla_{\o}^i\Phi'(\tilde{\lambda})\Ipr\nabla_{\o i}\Phi(\tilde{\lambda})\,.
\end{eqnarray}
We have organized the terms in a form corresponding to the types of physical effects they describe. 
Here IISW is a variation of ISW terms, including a single integral of a quadratic SW term and double and triple integrated terms.  Cross terms describe interactions, such as $\delta^2\NLd z_{\text{\tiny Dop}\p \times \text{\tiny ISW}\p }$ which describes an interaction between the line-of-sight Doppler and ISW terms.  
\end{description}

The final expression for the redshift up to second order is
\begin{equation}\label{eq:forredshiftexp}
1+\hat{z}_s = (1+\bar{z})\left[ 1 +\delta z +\frac{1}{2}\bigg\{\delta^2\Sd z+\delta^2\Vd z +\delta^2\Td z+ \delta^2\NLd z \bigg\}\right].
\end{equation}

\section{Area Distance as a function of affine parameter}\label{sec:area-ditance}

Here we present the area distance up to second order as a function of affine parameter~-- or equivalently, as a function of background redshift. We define the perturbations by 
\be
D_A(\lambda_s)=\bar D_A(\lambda_s)+\delta D_A(\lambda_s)+\frac{1}{2}\delta^2 D_A(\lambda_s)\,.
\ee
 This can be used with the formula \eqref{eq:forredshiftexp} for redshift up to second order as a function of affine parameter, giving the distance-redshift relation up to second order in parametric form.  

Under a conformal transformation~\eqref{conf3}, the area distance in the physical universe is $\hat D_A=a D_A$, i.e., 
\be
\hat D_A(\hat\lambda)=a(\lambda) D_A(\lambda), ~~~\d \lambda = a^{-2} \d \hat{\lambda}.
\ee 
This relationship simplifies most of the calculations because we do all calculations on a Minkowski background. 

We  calculate $D_A(\lambda)$ to second order by solving \eqref{CovAreadistance} with the boundary conditions \eqref{conf2}, using
\begin{equation}
\frac{\d^2 \delta D_A}{\d \lambda^2}=k^a\nabla_a\left(k^b\nabla_b \delta D_A\right)=k^ak^b\nabla_a\nabla_b \delta D_A\,.
\end{equation}
In general we can write \eqref{CovAreadistance} at any perturbative order $n$ in the form
\begin{equation}\label{eq:generaleqnforDA}
\frac{\d^2 \delta^{n} D_A}{\d \lambda^2}= \delta^{n} S(\lambda),
\end{equation}
where the source $\delta^{n}S(\lambda)$ contains the area distance up to order $n-1$, and perturbed metric variables. Using the boundary conditions
\be
\delta^{n} D_A=0~~~\mbox{and}~~~
\frac{\d\delta^n D_A}{\d\lambda}\bigg|_o = - \delta^n E_o\,,
\ee
where $\delta^nE_o$ is given in the previous section for $n=1,2$.
The general solution to (\ref{eq:generaleqnforDA}) is then
\begin{eqnarray}
\delta^{n} D_A(\lambda_s)
&=&(\lambda_o-\lambda_s)\delta^n E_o+\int^{\lambda_s}_{\lambda_o}(\lambda_s-\lambda) \delta^{n}S(\lambda)\d\lambda\,.
\end{eqnarray}

\subsection{Area distance in the background}


In the Minkowski background, (\ref{CovAreadistance}) simplifies to 
${\d^2D_A}/{\d\lambda^2}=0$,
since $R_{ab}=0$.
The  solution is $D_A= C_1 + \lambda C_2$. The boundary conditions \eqref{conf2} (where $E_o=1$) give $C_1=0$ and $C_2=-1$, so that
\be
D_A(\lambda_s)=(\lambda_o-\lambda_s)\,.
\ee 
In the FLRW background
 \begin{eqnarray}\label{DAeqn1}
 \hat D_A(\hat\lambda_s) 
 &=&a(\hat\lambda_s)(\hat\lambda_o-\hat\lambda_s)=a(\hat\lambda_s)\chi_s, \\ 
\hat D_A(\hat z_s)&=&  
 \frac{1}{(1+\hat{z}_s)}\int_{0}^{\hat{z}_s} \frac{\d z}{(1+z)\HH(z)}\,.
 \end{eqnarray}
 where $\hat{z}_s$ is the redshift corresponding to $\hat\lambda_s$.



\subsection{First-order contribution}

At first order \eqref{eq:generaleqnforDA} takes the form
 \begin{eqnarray}\label{eq:perteqarea}
\frac{\d^2 \delta D_A}{\d \lambda^2}&=& -\frac{1}{2}D_A \delta {R}_{ab}k^ak^b + D'_A{\d\delta k^0\over \d\lambda}+2D''_A\delta k^0
\,,
\end{eqnarray}
where we used $R_{ab}=0$ in the background. 
Substituting for $\delta k^0$ and $\delta R_{ab}$ (see Appendix A), \eqref{eq:perteqarea} becomes
    \begin{eqnarray}
  \frac{\d^2 \delta D_A}{\d \lambda^2}&=& -2\frac{\d D_A}{\d \lambda} n^i\D_i\Phi - D_A\left[\Phi''-2n^i\D_i\Phi'  +\D_i\D^i\Phi\right]\,\nonumber ,\\
  &=&2\frac{\d D_A}{\d\lambda}\left({\d_\lambda\Phi}-\Phi'\right)+D_A\left[\frac{2}{\lambda_o-\lambda}\left({\d_\lambda\Phi}-\Phi'\right)-{\d_\lambda^2\Phi}-\nabla_{\bot}^2\Phi\right]\,,\nonumber\\
  &=&-D_A\left[{\d_\lambda^2 \Phi}+\nabla_{\o}^2 \Phi\right]\,.\label{DAeqn1}
 \end{eqnarray}
In the second line, we have performed a full decomposition of double derivatives of $\Phi$ using (\ref{eq:spatialdecomposition1}) and have substituted for radial derivatives, while in the third line we simplified using the background solution for $D_A$.
The solution to  \eqref{DAeqn1} is, on performing several integrations by parts,
 \begin{eqnarray}\label{DAeqn1soln}
 \frac{\delta D_A }{\bar D_A}&=&  -\Phi_s+ \D\p  v_o- \frac{1}{(\lambda_o-\lambda_s)} \left\{ 2\I \Phi + \I {(\lambda_s-\lambda) ( \lambda_o- \lambda)}\nabla^2_{\bot}\Phi \right\}\, .
 \end{eqnarray}
This is proportional to the full relativistic weak lensing convergence, $\kappa$,  at first order, which includes SW, ISW and Doppler terms in addition to the standard gravitational lensing integral. Our result is in agreement  with  \cite{Bonvin:2008ni} (on noting the difference in the definition of $\nabla_\o^2$ used there).

\subsection{Area distance at second order}
  
At second  order \eqref{eq:generaleqnforDA} takes the form
  \begin{eqnarray}
  \frac{\d^2\delta^2 D_A}{\d\lambda^2}&=& -2\delta k^a\delta k^b \nabla_a\nabla_b D_A - 2k^b \delta^2 k^a \nabla_a \nabla_b D_A-4 \delta k^ak^b \nabla_a\nabla_b \delta D_A - 2 \delta k^a \nabla_a \delta k_b\nabla^b D_A \nonumber\\ &&
 -{\text{D}\delta^2 k_a\over \d\lambda} \nabla^a D_A- \delta^2k^a \nabla_a k_b\nabla^b D_A- 2 {\text{D}\delta k_a\over \d\lambda} \nabla^a \delta D_A - 2 \delta k^a \nabla_a k_b \nabla^b \delta D_A 
 -\delta D_A\delta R_{ab}k^ak^b
  \nonumber\\ &&
  +D_A\left[- \delta \Sigma_{ab}\delta \Sigma^{ab}-2\delta k^ak^b\delta R_{ab} -\frac{1}{2} \delta^2 R_{ab}k^ak^b \right] . \label{eq:areadistsecond1}
\end{eqnarray}
We split the various contributions on the right into induced and quadratic terms,  as previously.

  \begin{description}
  
  \item [\tt{Area distance from induced scalars}]
  
Considering only second-order induced scalars in \eqref{eq:areadistsecond1}:
  \begin{eqnarray}
  \frac{\d^2 \delta^2\Sd D_A}{\d\lambda^2}&=& \frac{\d D_A}{\d\lambda}\left( \Psi'\twod-\Phi'\twod+ 2\D\p \Phi\two\right)
  +D_A\left[\Psi''\twod-2\D\p \Psi'\twod+ \frac{1}{2}\D^2(\Phi\two+ \Psi\two)
 \right.  \nonumber\\ && \left.
  + \frac{1}{2}n^in^j\D_i\D_j\left(\Psi\two+ \Phi\two\right)\right].
  \end{eqnarray}
Decomposing  the spatial derivative further, so that the result may be expressed in terms of angular derivatives, we find 
\begin{eqnarray}
\frac{\d^2 \delta^2\Sd D_A}{\d\lambda^2} &=& \left( {\d\Psi\two\over \d\lambda}- {\d\Phi\two\over\d\lambda}\right) - D_A\left[{\d^2 \Psi\two\over \d\lambda^2}
+\frac{1}{2}\left( \nabla_{\bot}^2\Phi\two + \nabla_{\bot}^2\Psi\two\right) \right].
\end{eqnarray}
Following a similar procedure as in the first order case, we integrate both sides twice to obtain a solution for area distance with the effect of induced scalars:
\begin{eqnarray}
\frac{\delta^2\Sd D_{A}}{\bar D_A}&=& - \Psi\two_s+\D\p  v_o\two - \frac{1}{(\lambda_o-\lambda_s)} \I\left[
 (\Phi\two+ \Psi\two) 
 + \frac{1}{2}{(\lambda_s-\lambda)(\lambda_o-\lambda)} \nabla_{\bot}^2( \Phi\two+\Psi\two)\right]\,,
\end{eqnarray}
where we used (\ref{eq:deltasqEobsS}).

\item [\tt{ Area distance from induced vectors}]

For the contribution from induced vectors we have
\begin{eqnarray}\label{eq:secondorderVec1}
\frac{\d^2\delta^2\Vd D_A}{\d\lambda^2 }= \frac{\d D_A}{\d\lambda} n^in^j \D_i\omega_j + \frac{1}{2}D_A\left( n^in^j \D_i\omega'_j- n^i\D^2\omega_i\right)\,.
\end{eqnarray}
Some terms in   \eqref{eq:secondorderVec1} may easily be integrated if the spatial derivatives are decomposed further into screen space derivatives: 
\begin{eqnarray}\label{eq:secondorderVec2}
\frac{\d^2 \delta^2\Vd D_A}{\d\lambda^2} &=&  D_A\left[\frac{1}{(\lambda_o-\lambda)^2}\omega_{\pp} - \frac{1}{2}{\d^2 \omega_{\pp}\over \d\lambda^2} + \frac{1}{2}{\d \omega'_{\pp}\over \d\lambda}- \frac{1}{2}\nabla_{\bot}^2\omega_{\pp}\right]
\end{eqnarray}
Integrating both sides twice gives 
\begin{eqnarray}\label{eq:secondorderVec3}
\frac{\delta^2\Vd D_{A}}{\bar D_{A}}
&=& -\frac{1}{2}\omega_{\p s}+v_{\| o}\two + \frac{1}{2(\lambda_o-\lambda_s)} \I\bigg[2\frac{(\lambda_s-\lambda_o)}{(\lambda_o-\lambda)} \omega_{\pp}
+\left(\lambda_s+\lambda_o-2\lambda \right)\omega'_{\pp} -{ (\lambda_s-\lambda)(\lambda_o-\lambda)}\nabla_{\bot}^2\omega_{\pp} \bigg],
\end{eqnarray}
where we used (\ref{eq:deltasqEobsV}).

\item [\tt{ Area distance from induced tensors}]

For tensors, \eqref{eq:areadistsecond1} gives 
\begin{eqnarray}\label{secondorderTensor}
\frac{\d^2 \delta^2\Td D_A}{\d\lambda^2 } &=& \frac{\d D_A}{\d\lambda} h'_{ij} n^in^j + \frac{1}{2} D_A\left[ h_{ij}'' - \nabla^2h_{ij}\right]n^in^j.
\end{eqnarray}
Decomposing the 3-tensor $h_{ij}$  into 2-scalars, 2-vectors and an irreducible 2-tensor on the screen space, as in \eqref{3t2dec},
Then
\begin{eqnarray}\label{secondorderTensor2}
\frac{\d^2 \delta^2\Td D_A}{\d\lambda^2 } &=& -{\d h{\p}\over \d\lambda}+ D_A\left[ -\frac{3}{(\lambda_o-\lambda)^2} h{\p}  - {\d h'{\p} \over \d\lambda} +\frac{1}{2}{\d^2 h{\p}\over \d \lambda^2}+ \frac{1}{2} \nabla_{\bot}^2 h_{\pp}\right]\,.
\end{eqnarray}
Using  $\delta^2\Td E_o = 0$, the solution becomes
\begin{eqnarray}
\frac{\delta^2\Td D_A}{\bar D_A}&=& \frac{1}{2}( h_{\pp s} -h_{\pp o}) - \frac{1}{(\lambda_o-\lambda_s)}\I\bigg[ 3\frac{(\lambda_s-\lambda)}{(\lambda_o-\lambda)} h{\p} 
+ \left(\lambda_s+\lambda_o-2\lambda \right)h'{\p} 
 - \frac{1}{2} {(\lambda_s-\lambda)(\lambda_o-\lambda)} \nabla_{\bot}^2 h{\p}\bigg]\,.
\end{eqnarray}

\item [\tt{ Area distance from quadratic terms}]

The contribution to area distance
from terms quadratic in first order perturbations is more complicated.
First we decompose the 4-covariant derivative in  \eqref{eq:areadistsecond1} into temporal and spatial parts and substitute  for the Ricci tensor using  \eqref{eq:RicciNonlinear}: 
\begin{eqnarray}
\frac{\d^2 \delta^2\NLd D_A}{\d\lambda^2} &=&4\nabla_{\pp}\delta D_A{\d\Phi\over \d\lambda}
+4{\d \delta D_A\over\d\lambda}\left({\d\Phi\over\d\lambda}-\Phi'\right)
+8\nabla_{\pp}{\d\delta D_A\over\d\lambda} (\Phi-\Phi_o)
+8\nabla_{\o i}{\d\delta D_A\over\d\lambda} \I\nabla_{\o }^i\Phi \nonumber\\ &&
+ 4 \nabla_{\o i}\Phi \nabla_{\o}^i\delta D_A+ 2\delta D_A\left[2{\chi}^{-1}\left({\d\Phi\over\d\lambda}-\Phi'\right)-{\d^2\Phi\over\d\lambda^2}-\nabla^2_{\o}\Phi\right]
 \nonumber\\ &&
+D_A\bigg\{-\delta \Sigma_{ij}\delta \Sigma^{ij}-2\bigg({\d\Phi\over\d\lambda}\bigg)^2+ 8 \nabla_{\o i}\Phi'\I\nabla^i_{\o}\Phi+ 4 \Phi\bigg(2{ \d\Phi'\over \d\lambda}-3{\d^2\Phi\over\d\lambda^2}-2 \nabla_{\o}^2\Phi\bigg)
 \nonumber\\ &&
+ 4 \Phi_{o}\left({\d^2\Phi\over\d\lambda^2} -2 \Phi'' + \nabla^2_{\o}\Phi\right)
+\frac{8}{(\lambda_o-\lambda)}\left[\Phi{\d\Phi\over\d\lambda}-\Phi'(\Phi-\Phi_o)
+\nabla_{\o i}\Phi\I\nabla_{\o}^i \Phi \right] \label{eq:NL1}
\bigg\}
\end{eqnarray}
At second order, the shear contributes to the nonlinear component of  the area  distance as a square of  contribution from the first order scalar perturbations. We calculate this from (\ref{eq:shearevo}), 
 \begin{eqnarray}
\frac{\d \delta  {\Sigma}_{ij}}{\d {\lambda}}
=-2\left( N_i{}^k N_j{}^l-\frac{1}{2}N_{ij}N^{kl}\right)\D_k\D_l\Phi\,.
\end{eqnarray}
Without loss of generality, we set the perturbation of shear at the observer $\delta \hat{\Sigma}^{ab}{}_o=0$, then we have
\begin{eqnarray}\label{shear2}
\delta {\Sigma}_{ij} &=&-2\I \nabla_{\<i} \nabla_{ j\>}\Phi\,.
\end{eqnarray}
We also require various derivatives of the first-order area distance, which are:
With the use of 
\begin{eqnarray}
\delta D_A(\lambda) &=& -(\lambda_o-\lambda)\left(\Phi-\D_{\pp}v_o\right)
-2\I\Phi-\I(\lambda_s-\lambda)(\lambda_o-\lambda)\nabla_{\o}^2 \Phi \,,\\
\nabla_{\o i}\delta D_A&=& -(\lambda_o-\lambda)\nabla_{\o i}\Phi -2 \I\frac{(\lambda_o-\lambda)}{(\lambda_o-\lambda_s)}\nabla_{\o i} \Phi - \I \frac{(\lambda_s-\lambda)(\lambda_o-\lambda)^2}{(\lambda_o-\lambda_s)}\nabla_{\o i}\nabla_{\o}^2\Phi\,,\\
\frac{\d\delta D_A}{\d\lambda}&=&-(\Phi+\D_{\pp}v_o)-(\lambda_o-\lambda)\left(\Phi'-\D_{\pp}\Phi\right)-\I(\lambda_o-\lambda)\nabla_{\o}^2\Phi\,,\\
\nabla_{\o i}\frac{\d\delta D_A}{\d\lambda}&=& -\nabla_{\o i}\Phi- (\lambda_o-\lambda)\left(\nabla_{\o i}\Phi'-\D_{\pp}\nabla_{\o i}\Phi\right)-\I\frac{(\lambda_o-\lambda)^2}{(\lambda_o-\lambda_s)}\nabla_{\o i}\nabla_{\o}^2 \Phi \,,\\
\D_{\pp}\delta D_A&=& -(\lambda_o-\lambda)\D_{\pp}\Phi+ 3\Phi-\D_{\pp}v_o+ \I(\lambda_o-\lambda)\nabla_{\o}^2\Phi\,,\\
\D_{\pp}\frac{\d\delta D_A}{\d\lambda}&=&-\Phi'-(\lambda_o-\lambda)\left(\D_{\pp}\Phi'-\D_{\pp}^2\Phi+\nabla_{\o} ^2\Phi\right)\,.
\end{eqnarray}

Substituting these expressions, as well as \eqref{shear2},  into \eqref{eq:NL1}  and simplifying
spatial derivatives by  decomposing into derivatives on the screen space and a  derivative along the line of sight we have:
\begin{eqnarray}
\frac{\d^2 \delta^2\NLd D_A}{\d\lambda^2} &=&4\Phi'\left[\D_{\pp}v_o-3\Phi+\frac{2}{(\lambda_o-\lambda)}\I\Phi+\I(\lambda_o-\lambda)\nabla_{\o}^2\Phi\right]
-4\nabla_{\o i}\Phi\left[2\I\frac{(\lambda_o-\lambda)}{(\lambda_o-\lambda_s)}\nabla_{\o}^i\Phi
\right.  \nonumber\\ &&\left.
+\I\frac{(\lambda_s-\lambda)(\lambda_o-\lambda)^2}{(\lambda-\lambda_s)^2}\nabla_{\o}^i\nabla_{\o}^2\Phi\right]
+\frac{4}{(\lambda_o-\lambda)}\Phi\I(\lambda_s-\lambda)(\lambda_o-\lambda)\nabla_{\o}^2\Phi
-8\I\nabla_{\o i}\Phi\I\frac{(\lambda_o-\lambda)^2}{(\lambda_o-\lambda_s)}\nabla_{\o}^i\nabla_{\o}^2\Phi
 \nonumber\\ &&
+4{\d}_{\lambda}\Phi\left[2\D_{\pp}v_o-\frac{1}{(\lambda_o-\lambda)}\I\Phi-\frac{1}{(\lambda_o-\lambda)}\I(\lambda_s-\lambda)(\lambda_o-\lambda)\nabla_{\o}^2\Phi+6 \Phi\right]
+2{\d^2_\lambda}\Phi\left[2\I\Phi
\right.  \nonumber\\ &&\left.
+\I(\lambda_s-\lambda)(\lambda_o-\lambda)\nabla_{\o}^2\Phi\right]
+\bar{D}_A\bigg\{-4\Phi_o\left[2\Phi''-3\nabla_{\o}^2\Phi\right]
-2\D_{\pp}v_o\left[\frac{1}{(\lambda_o-\lambda)}\Phi'+\nabla^2_{\o}\Phi\right]
 \nonumber	\\ &&
+2\Phi
\left[\frac{2}{(\lambda_o-\lambda)}\Phi'- 7 \nabla_{\o}^2\Phi\right]
-8\nabla_{\o i}\Phi'\I\nabla^i_{\o}\Phi
-4\nabla_{\o i}\Phi\nabla_{\o}^i\Phi
-4\I\nabla_{\o\<i}\nabla_{\o j\>}\Phi\I\nabla_{\o}^{\<i}\nabla_{\o}^{ j\>}\Phi\bigg]
\nonumber \\	&&
-2\left(\d_{\lambda}\Phi\right)^2
-4\Phi_o\left[{\d^2_\lambda}\Phi-2{\d}_{\lambda}\Phi'\right]
-2\D_{\pp}v_o\left[{\d^2_\lambda\Phi}-\frac{2}{(\lambda_o-\lambda)}{\d}_{\lambda}\Phi\right]
 \nonumber\\ &&
-2\Phi\left[\frac{2}{(\lambda_o-\lambda)}+{\d^2_\lambda}\Phi\right]
-8\nabla_{\o i}{\d}_{\lambda}\Phi\I\nabla^i_{\o}\Phi
 \bigg\}\,. \label{eq:NL2}
\end{eqnarray}
Integrating (\ref{eq:NL1}) and simplifying we have 
\begin{eqnarray}
\frac{\delta^{2}\NLd D_A}{D_A}
&=&-\Phi^2_o
  -4 \Phi_o  \D_{\| }v_o
  +\D_{\pp}v_o \D_{\pp} v_o+\nabla_{\o i}v_o\nabla_{\o}^iv_o
  +\I\frac{(\lambda_s-\lambda)}{(\lambda_o-\lambda_s)}
  \bigg\{
  4\Phi'\left[\D_{\pp}v_o-3\Phi
  +\frac{2}{(\lambda_o-\lambda)}\I\Phi
\right.  \nonumber\\ && \left.    
  +\I(\lambda_o-\lambda)\nabla_{\o}^2\Phi\right]
-4\nabla_{\o i}\Phi\left[2\I\frac{(\lambda_o-\lambda)}{(\lambda_o-\lambda_s)}\nabla_{\o}^i\Phi+\I\frac{(\lambda_s-\lambda)(\lambda_o-\lambda)^2}{(\lambda-\lambda_s)^2}\nabla_{\o}^i\nabla_{\o}^2\Phi\right]
\nonumber\\ &&
+\frac{4}{(\lambda_o-\lambda)}\Phi\I(\lambda_s-\lambda)(\lambda_o-\lambda)\nabla_{\o}^2\Phi
-8\I\nabla_{\o i}\Phi\I\frac{(\lambda_o-\lambda)^2}{(\lambda_o-\lambda_s)}\nabla_{\o}^i\nabla_{\o}^2\Phi
\bigg\}
   \nonumber\\ &&
  +\I\frac{(\lambda_s-\lambda)(\lambda_o-\lambda)}{(\lambda_o-\lambda_s)}
  \bigg\{-4\Phi_o\left[2\Phi''-3\nabla_{\o}^2\Phi\right]
-2\D_{\pp}v_o\left[\frac{1}{(\lambda_o-\lambda)}\Phi'+\nabla^2_{\o}\Phi\right]
 \nonumber \\	&&
+2\Phi\left[\frac{2}{(\lambda_o-\lambda)}\Phi'-7\nabla_{\o}^2\Phi\right]
-8\nabla_{\o i}\Phi'\I\nabla^i_{\o}\Phi
-4\nabla_{\o i}\Phi\nabla_{\o}^i\Phi
-4\I\nabla_{\o\<i}\nabla_{\o j\>}\Phi\I\nabla_{\o}^{\<i}\nabla_{\o}^{ j\>}\Phi \bigg\}
\nonumber \\	&&
+\I\frac{(\lambda_s-\lambda)}{(\lambda_o-\lambda_s)}
  \bigg\{4{\d}_{\lambda}\Phi\left[2\D_{\pp}v_O
-\frac{1}{(\lambda_o-\lambda)}\I\Phi-\frac{1}{(\lambda_o-\lambda)}\I(\lambda_s-\lambda)(\lambda_o-\lambda)\nabla_{\o}^2\Phi+6 \Phi\right]
 \nonumber \\&&
+2{\d}^2_{\lambda}\Phi\left[\I(\lambda_s-\lambda)(\lambda_o-\lambda)\nabla_{\o}^2\Phi
+2\I\Phi\right] \bigg\}
+\I\frac{(\lambda_s-\lambda)(\lambda_o-\lambda)}{(\lambda_o-\lambda_s)}
  \bigg\{-2\left({\d}_{\lambda}\Phi\right)^2
-4\Phi_o\left[{\d}^2_{\lambda}\Phi-2{\d}_{\lambda}\Phi'\right]  \nonumber\\ &&
-2\D_{\pp}v_o\left[{\d}^2_{\lambda}\Phi-\frac{2}{(\lambda_o-\lambda)}{\d}_{\lambda}\Phi\right]
-2\Phi\left[\frac{2}{(\lambda_o-\lambda)}+{\d}^2_{\lambda}\Phi\right]
-8\nabla_{\o i}{\d}_{\lambda}\Phi\I\nabla^i_{\o}\Phi
 \bigg\}\,.
 \label{eq:quadsoln1}
  \end{eqnarray}

Upon several integration by parts in the last three lines of \eqref{eq:quadsoln1}, we have the alternative representation:
\begin{eqnarray}
\frac{\delta^{2}\NLd D_A}{D_A}
&=&-\Phi_o\left(4\Phi_s-6\Phi_o\right)+2\D_{\pp} v_o\left(\Phi_s-7\Phi_o\right)+\D_{\pp}v_o \D_{\pp} v_o+
  \nabla_{\o i}v_o\nabla_{\o}^iv_o
  +\frac{4}{\chi_s}\left(\Phi_s-2\Phi_o\right)\I\Phi
\nonumber \\ \nonumber&&  
  +4\Phi_o\left[2\I\frac{(\lambda_o-\lambda)}{(\lambda_o-\lambda_s)}\Phi'+2\I\frac{(\lambda_s-\lambda)}{(\lambda_o-\lambda_s)}
  \Phi'
  -2\I\frac{(\lambda_s-\lambda)(\lambda_o-\lambda)}{(\lambda_o-\lambda_s)}\Phi''\right]
  -16\I\frac{(\lambda_s-\lambda)}{(\lambda_o-\lambda_s)}\Phi\Phi'
  \\ \nonumber&&  
  +\frac{2}{(\lambda_o-\lambda_s)}\I\Phi^2  
  + 2\left(6\Phi_o+\Phi_s\right)\I\frac{(\lambda_s-\lambda)(\lambda_o-\lambda)}{(\lambda_o-\lambda_s)}\nabla^2_{\o}\Phi
  +2\D_{\pp}v_o\left[\frac{4}{(\lambda_o-\lambda_s)}\I\Phi
\right.   \\ \nonumber&&\left.  
  -\I\frac{(\lambda_s-\lambda)(\lambda_o-\lambda)}{(\lambda_o-\lambda_s)}\nabla_{\o}^2\Phi\right]  
  -4\I\frac{1}{(\lambda_o-\lambda)(\lambda_o-\lambda_s)}\Phi\Ipr\Phi(\tilde{\lambda})
  -4\I\frac{(\lambda_s-\lambda)}{(\lambda_o-\lambda)(\lambda_o-\lambda_s)}
  \\ \nonumber&&\times  
  \left[\frac{1}{(\lambda_o-\lambda)}\Phi\Ipr\Phi(\tilde{\lambda})
  -\Phi^2-2\Phi'\Ipr\Phi(\tilde{\lambda})\right]
  -4\I\frac{1}{(\lambda_o-\lambda)}\Phi\Ipr\frac{(\lambda-\tilde{\lambda})(\lambda_o-\tilde{\lambda})}{(\lambda_o-\lambda_s)}\nabla_{\o}^2\Phi(\tilde{\lambda})
     \\ \nonumber&&
  -4\I\frac{(\lambda_s-\lambda)}{(\lambda_o-\lambda)^2(\lambda_o-\lambda_s)}\Phi\Ipr{(\lambda-\tilde{\lambda})(\lambda_o-\tilde{\lambda})}\nabla_{\o}^2\Phi(\tilde{\lambda})
  -4\I\frac{(\lambda_s-\lambda)^2}{(\lambda_o-\lambda_s)}\Phi\nabla^2_{\o}\Phi
    \\ \nonumber&&
  -4\I\Phi\Ipr\frac{(\lambda_o-\tilde{\lambda})}{(\lambda_o-\lambda_s)}\nabla_{\o}^2\Phi(\tilde{\lambda})
  -14\I\frac{(\lambda_s-\lambda)(\lambda_o-\lambda)}{(\lambda_o-\lambda_s)}\Phi\nabla_{\o}^2\Phi
  \\ \nonumber&&
  +4\I\frac{(\lambda_s-\lambda)}{(\lambda_o-\lambda_s)}\nabla_{\o}^2\Phi\Ipr\Phi(\tilde{\lambda})
  +4\I\frac{(\lambda_s-\lambda)}{(\lambda_o-\lambda_s)}\Phi'\Ipr(\lambda_o-\tilde{\lambda})\nabla_{\o}^2\Phi(\tilde{\lambda})
  \\ \nonumber&&
  +4\I\frac{(\lambda_s-\lambda)}{(\lambda_o-\lambda)}\Phi\Ipr\frac{(\lambda-\tilde{\lambda})(\lambda_o-\tilde{\lambda})}{(\lambda_o-\lambda_s)}\nabla_{\o}^2\Phi(\tilde{\lambda})
  -8\I\frac{(\lambda_s-\lambda)}{(\lambda_o-\lambda_s)}\nabla_{\o i}\Phi\Ipr\frac{(\lambda_o-\tilde{\lambda})}{(\lambda_o-\lambda_)}\nabla_{\o}^i\Phi(\tilde{\lambda})
\\ \nonumber&&  
  -8\I\frac{(\lambda_o-\lambda)}{(\lambda_o-\lambda_s)}\nabla_{\o i}\Phi\Ipr\nabla^i_{\o}\Phi(\tilde{\lambda})
  -\frac{8}{(\lambda_o-\lambda_s)}\I\Ipr\nabla_{\o i}\Phi(\tilde{\lambda})\Ipr\nabla_{\o}^i\Phi(\tilde{\lambda})
\\ \nonumber&&  
  -12\I\frac{(\lambda_s-\lambda)(\lambda_o-\lambda)}{(\lambda_o-\lambda_s)}\nabla_{\o i}\Phi\nabla_{\o}^i\Phi
  -8\I\frac{(\lambda_s-\lambda)(\lambda_o-\lambda)}{(\lambda_o-\lambda_s)}\nabla_{\o i}\Phi'\Ipr\nabla_{\o}^i\Phi(\tilde{\lambda})
  \\ \nonumber&&
  -2\I\frac{(\lambda_s-\lambda)}{(\lambda_o-\lambda_s)}\nabla_{\o}^2\Phi\Ipr(\lambda-\tilde{\lambda})(\lambda_o-\tilde{\lambda})\nabla_{\o}^2\Phi(\tilde{\lambda})
  -8\I\frac{(\lambda_s-\lambda)}{(\lambda_o-\lambda_s)}\Ipr\nabla_{\o i}\Phi(\tilde{\lambda})\Ipr\frac{(\lambda_o-\tilde{\lambda})^2}{(\lambda_o-\lambda)}\nabla_{\o}^i\nabla^2\Phi(\tilde{\lambda})
  \\ \nonumber&&
  -4\I\frac{(\lambda_s-\lambda)(\lambda_o-\lambda)}{(\lambda_o-\lambda_s)}\Ipr\nabla_{\o\<i}\nabla_{\o j\>}\Phi(\tilde{\lambda})\Ipr\nabla_{\o}^{\<i}\nabla_{\o}^{ j\>}\Phi(\tilde{\lambda})
  \\ &&
  -4\I\frac{(\lambda_s-\lambda)}{(\lambda_o-\lambda_s)}\nabla_{\o i}\Phi\Ipr\frac{(\lambda_s-\tilde{\lambda})(\lambda_o-\tilde{\lambda})^2}{(\lambda_o-\lambda)}\nabla_{\o}^i\nabla_{\o}^2\Phi(\tilde{\lambda})\,.
  \end{eqnarray}
 Thus the area distance as a function of affine parameter in a perturbed FLRW spacetime is given by 
\begin{eqnarray}\label{Areadistance}\label{Areadistance2}
 \hat D_A(\lambda_s)&=& a(\lambda_s)(\lambda_o-\lambda_s)\left[1+ \frac{\delta D_A }{{\bar D}_A}(\lambda_s)+ \frac{1}{2}\frac{\delta^2 D_A}{{\bar D}_A}(\lambda_s)\right],\\
\delta^2 D_A&=&\delta\Sd^2 D_A+\delta\Vd^2 D_A+\delta\Td^2 D_A+\delta\NLd^2 D_A\,.
\end{eqnarray}
This is given at an affine parameter distance to the source $\lambda_s$ corresponding to redshift $\bar z_s$ in the background. 
  \end{description}
 
 We now have the physical area distance $\hat D_A(\hat z)$, given as a parametric relation between $\hat D_A(\lambda)$ and $\hat z(\lambda)$~\eqref{eq:reshiftexp}, using the affine parameter as a parametric parameter. This is suitable for plotting $\hat D_A(\hat z)$ along a line of sight where the potential is known, as in ray tracing in an N-body simulation, but is not suitable where we are given redshift information directly. We turn now to the final steps to compute $\hat D_A(\hat z)$ explicitly.

\section{Area distance as a function of observed redshift}\label{sec:obs}
   
To calculate $\hat D_A(\hat z)$ explicitly we need to perturbatively invert $\hat z(\lambda)$ for $\lambda(\hat z)$ and substitute this into \eqref{Areadistance2}. That is, we need $\hat D_A$ on surfaces of constant $\hat z$ (the observed redshift) rather than on surfaces of constant $\lambda$ (which is not observable).  This corresponds to a gauge change from the $\lambda$ coordinate on the past null cone to $\hat z$. Each $\hat z$~-- even though it is observed in the perturbed space-time~-- is a number which can be mapped to a background scale factor and affine parameter. From this we develop our perturbative expansion for $\lambda$. Since the background function $z(\lambda)$ relation is monotonic, any $\hat z$ value has a corresponding $\lambda$ associated with it.\footnote{This is analogous to switching from coordinate time in the Poisson gauge to proper time in the comoving synchronous gauge. If we have the density $\rho(t)$ to second-order, and the proper time along comoving worldlines $\tau(t)$, 
also to second-order, we can eliminate the variable $t$ to find $\rho(\tau)$ on constant proper-time hypersurfaces perturbatively as follows. Write $t=t_0+\delta t+\frac{1}{2}\delta^2 t$ and invert $\tau(t)$ order by order, and substitute for $t$ in $\rho(t)$ to give $\rho(\tau)$ explicitly. Then there are terms in this such as $\delta\tau(t)$ which become $\delta\tau(t_0)$ when they are to be evaluated. This change gives corrections to the first- and second-order density, reflecting the different density perturbations in the different gauges.}     
   
To obtain the observed area distance explicitly as a function of redshift, we expand the `background' $\lambda$ as
  \begin{equation}
  {\lambda}= {\lambdaz} +\delta \lambda + \frac{1}{2}\delta^2 \lambda
  \end{equation}
where $\lambdaz$ is the affine parameter in redshift space corresponding to the redshift $\hat z$ \emph{as if there were no perturbations}. In other words, we define $\lambdaz$ using the background relation
\be\label{z-def}
\frac{1}{a({\lambdaz})}=1+\hat z\,.
\ee
This relation serves to anchor $\delta\lambda$ and $\delta^2\lambda$ as follows. This value of $\lambdaz$ has associated derivatives of $a$ at any $\hat z$:
  \begin{eqnarray}
 \frac{1}{a}\frac{\d a}{\d \lambdaz}&=&  \HH(\lambdaz), \\ 
 \frac{1}{a}\frac{\d^2 a}{\d \lambdaz^2}&=&\left[\frac{\d \HH (\lambdaz)}{\d \lambdaz} + \HH^2(\lambdaz)\right]\,.
 \end{eqnarray}
Expanding the scale factor about $\lambda$, we have 
 \begin{eqnarray}\label{perta}
 a(\lambda) 
&=& a({\lambdaz})\left[1+ \HH\delta \lambda+\frac{1}{2}\HH\,{\delta}^{2}\lambda  +\frac{1}{2}\left[ \frac{\d \HH }{\d \lambda }+ \HH^2\right]  \delta \lambda^2+\mathcal{O}(\delta^3\lambda)\right]\,.
 \end{eqnarray} 

Inverting~\eqref{eq:forredshiftexp} and  then expanding in  power series, the resulting equation becomes, after  using  (\ref{perta}),
   \begin{eqnarray}\label{phya}
 \frac{1}{(1+\hat{z} )}&=& \frac{a({\lambdaz})}{a({\lambdaz}_o)}\left[1+\left(\HH \delta \lambda - \delta z\right)+ \left(\frac{1}{2}\HH  \delta^2 \lambda- \frac{1}{2}\delta^2 z + \delta z^2 - \HH  \delta z \delta \lambda 
 + \frac{1}{2}\left(\frac{\d \HH  }{\d \lambdaz} + \HH^2 \right)\delta \lambda^2\right)+\mathcal{O}(\delta^3\lambda)\right]\,.
 \end{eqnarray}
Then for consistency with \eqref{z-def} we find
 \begin{eqnarray}
 \delta \lambda &=&\frac{\delta z}{  \HH}, \\
  \delta^2\lambda&=& \frac{1}{\HH}\left[\delta^2 z -(\delta z)^2 \left(1+ \frac{1}{\HH^2}\frac{\d\HH}{\d\lambdaz}\right)\right]\,.\label{eq;consistency}
 \end{eqnarray}
 Note that $\d\HH/\d\lambdaz=\HH'$ when multiplying a second-order quantity.
Using these relations to substitute for $a(\lambda)(\lambda_o-\lambda)$, we find that the area distance~\eqref{Areadistance2} becomes 
\begin{eqnarray}\label{DAgen2}
\nonumber
\hat D_A(\lambdaz)&=&a(\lambdaz )(\lambdaz_o -\lambdaz ) \left\{1+\left[\frac{\delta D_A}{\bar D_A}(\lambda) +\left(1- \frac{1}{\HH  (\lambdaz_o -\lambdaz )}\right)\delta z(\lambda)
\right]
 +\frac{1}{2} \left[ 
 \frac{ \delta^2 D_A}{\bar D_A}+
\left(1- \frac{1}{\HH  (\lambdaz_o -\lambdaz )}\right) \delta^2 z
\right.\right.\\ && \left.\left.
+\frac{\HH'-\HH^2}{\HH^3(\nu_o-\nu)} \delta z^2
+2\left(1- \frac{1}{\HH  (\lambdaz_o -\lambdaz )}\right)\frac{\delta D_A}{\bar D_A} \delta z
 \right]
 \right\}.
 \end{eqnarray}
 
 So far we have converted the scale factor in \eqref{Areadistance2} from $\lambda$ to $\lambdaz$. The first-order contribution in this expression also needs converting, since this gives additional second-order contributions (as they do not vanish in the background). For a general first-order quantity $\delta X(\lambda)$, converting to $\lambdaz$ gives
\be
\delta X(\lambda)=\delta X(\lambdaz)+\partial_\lambda \delta X\big|_\lambdaz \frac{\delta z(\lambdaz)}{\HH(\lambdaz)},
\ee  
 where $\delta X(\lambdaz)$ is understood to be $\delta X(\lambda\mapsto\lambdaz)$. 
 For $\partial_\lambda \delta X\big|_\lambdaz$ we are multiplying it by a first-order quantity so this is understood to be a derivative acting on the background. Hence we can just write $\d_\lambdaz \delta X$. 
 
With this, the area distance~\eqref{DAgen2} finally becomes 
 \begin{eqnarray}
\hat D_A(\hat z)&=&\frac{\nu_o-\nu}{1+\hat z} \left\{1+\left[\frac{\delta D_A}{\bar D_A}+\left(1- \frac{1}{\HH (\nu_o-\nu)}\right)\delta z\right]
 +
 \frac{1}{2}\left[\frac{ \delta^2 D_A}{\bar D_A} +\left(1- \frac{1}{\HH (\nu_o-\nu)}\right)\delta^2 z
  \right. \right. \nonumber\\ &&\left.\left.
 + 2 \left(1- \frac{1}{\HH (\nu_o-\nu)}\right)\left(\frac{\delta D_A}{\bar D_A}+\frac{\d_{\nu}\delta z}{\HH}\right)\delta z+2\d_{\nu}\frac{\delta D_A}{\bar D_A} \frac{\delta z}{\HH}
 + \frac{\HH'-\HH^2}{\HH^3(\nu-\nu_o)} \delta z^2\right]
 \right\}.\label{DAgen3}
 \end{eqnarray}
In this equation it is now understood that for functions such as $\delta z$, they are evaluated at $\lambdaz$ rather than $\lambda$. For example, 
\begin{eqnarray}\label{eq:redshift1}
{\delta z}(\lambdaz)
&=&\left[\D\p v(\lambdaz)-\Phi(\lambdaz)\right]-\left[\D\p v(\lambdaz_o)-\Phi(\lambdaz_o)\right]+2\int_{\lambdaz_o}^\lambdaz\Phi'd\lambdaz
=\left(\D\p v-\Phi\right)\big|^s_o +2\Iz  \Phi'\,,
\end{eqnarray}
where in the last equality we keep the abbreviated notation from above, but indicate the different integration variable.
Similarly, for derivatives of perturbed quantities we have explicitly  
\begin{eqnarray}
\d_{\lambdaz}\delta z(\lambdaz)
&=&\D{\p}v'-\D\p^2v+\D{\p}\Phi+\Phi'\,,
\\
\d_{\lambdaz_s}\left[\frac{\delta D_A}{D_A}(\lambdaz_s)\right] 
&=&
-\left(\Phi'-\D_{\pp}\Phi\right)+\frac{1}{(\nu_o-\nu_s)^2}\bigg[\Iz\left(2\Phi+(\nu_s-\nu)(\nu_o-\nu)\nabla_{\o}^2\Phi\right)\bigg]
\nonumber\\ &&
-\frac{1}{(\nu_o-\nu_s)}\left[2\Phi+\Iz(\nu_o-\nu)\nabla_{\o}^2\Phi\right]\,.
\label{eq:born2}
\end{eqnarray}
All contributions from converting $\lambda\to\lambdaz$ have now been taken into account, so that $\lambdaz$ \emph{is now treated as a background variable again}.  We now have $\hat D_A$ explicitly as a function of redshift. Although this is written as integrals over $\lambdaz$, this can be trivially converted to $\hat z$ using~\eqref{z-def}, or to $\chi_s=\nu_o-\nu_s$, or any other background variable of choice. 

Combining all the results together we can now give our final result, which we present using the background comoving distance $\chi_s(\hat z_s)$ -- which in turn is the distance to the source corresponding to the observed redshift $\hat z_s$ calculated from~\eqref{DAeqn1}: 
 
 \begin{eqnarray}
\hat D_A&=&\frac{\chi_s }{(1+\hat{z}_s)}\bigg\{1+\left(1- \frac{1}{\HH_s \chi_s}\right)\Phi_o
-\left(2- \frac{1}{\HH_s \chi_s}\right)\Phi_s+\frac{1}{\HH_s \chi_s}\D_{\pp}v_o
+\left(1- \frac{1}{\HH_s \chi_s}\right)\D_{\pp}v_s-2\left(1- \frac{1}{\HH_s \chi}\right)\Ic  \Phi'
\nonumber \\ \nonumber&&
+ \frac{1}{\chi_s} \left\{ 2\Ic \Phi 
+ \Ic {(\chi-\chi_s) \chi}\nabla^2_{\bot}\Phi \right\}
+\frac{1}{2}\bigg[\left(1- \frac{1}{\HH_s \chi_s}\right)\Phi\two_o+\frac{1}{\HH_s \chi_s}\D\p  v_o\two+\left(1- \frac{1}{\HH_s \chi_s}\right)\D\p  v\two_s- \Psi\two_s
\\ \nonumber&&
-\left(1- \frac{1}{\HH_s \chi_s}\right)\left(\Phi\two_s+\Ic\left( \Phi\twod' + \Psi\twod'\right)\right)
+ \frac{1}{\chi_s} \Ic(\Phi\two+ \Psi\two) 
 +\frac{1}{2}\Ic \frac{(\chi-\chi_s)\chi}{\chi_s} \nabla_{\bot}^2( \Phi\two+\Psi\two)
\bigg]\\ \nonumber&&
+\frac{1}{2}\bigg[-\frac{1}{\HH_s\chi_s}v_{\p o}\two-\left(1- \frac{1}{\HH_s \chi_s}\right)\omega_{\p o}+\left(1- \frac{1}{\HH_s \chi_s}\right)v_{\p s}\two
+\frac{1}{2}\left(1- \frac{2}{\HH_s \chi_s}\right)\omega_{\p s}
\\ \nonumber&& 
-\Ic\left(1- \frac{1}{\HH_s \chi_s}+\frac{\left(2\chi-\chi_s\right)}{2\chi_s}\right){\omega\p}'
-\frac{1}{\chi_s} \Ic\left(\frac{(\chi-\chi_s)}{\chi} \omega_{\pp} 
-\frac{1}{2}{ (\chi-\chi_s)\chi}\nabla_{\bot}^2\omega_{\pp}\right) \bigg]\\ \nonumber&&
+\frac{1}{2}\bigg[-\frac{1}{2} h_{\p s}+\Ic \left(1- \frac{1}{\HH_s \chi_s}-\frac{\left(2\chi-\chi_s \right)}{\chi_s}\right){h\p}'+ \frac{1}{\chi_s}\Ic\bigg( 3\frac{(\chi-\chi_s)}{\chi} h_{\pp} 
 - \frac{1}{2} {(\chi-\chi_s)\chi} \nabla_{\bot}^2 h_{\pp}\bigg)\bigg]\\ \nonumber&&
 +\frac{1}{2}\bigg\{\left(\frac{\HH'}{\HH^2}-1\right)\left[\left(\D_{\pp}v_s-\D_{\pp}v_o\right)^2
 -\left(\Phi-\Phi_o\right)^2\right]
 +2\left(3-\frac{1}{\chi_s\HH_s}\right)\Phi_s(\Phi_s-\Phi_o)  
 +6\Phi_o^2
 +2\chi(\Phi'_s-\D_{\pp}\Phi)
 \\ \nonumber&& \times
 \left[(\Phi-\Phi_o)
 -\left(\D_{\pp}v_s-\D_{\pp}v_o\right)\right] 
 +\frac{1}{\chi_s\HH_s}\nabla_{\o i}v_o\nabla_{\o}^iv_o+2\D_{\pp}v_o(\Phi_s-7\Phi_o)
 -2\left[\left(\frac{\HH'}{\HH^2}+1\right)\Phi_s-\left(\frac{\HH'}{\HH^2}-1\right)\Phi_o\right]
\\ \nonumber&&\times 
 \left(\D_{\pp}v_s-\D_{\pp}v_o\right)
+\D_{\pp}v_o \D_{\pp} v_o
+\left(1-\frac{1}{\chi_s\HH_s}\right)
\bigg[2(\D_{\pp}v_s)^2
+\nabla_{\o i}v_s\nabla_{\o}^iv_s-4\Phi_s\D_{\pp}v_s+\Phi_o\D_{\pp}v_o
-2\D_{\pp}v_o\D_{\pp}v_s
\\ \nonumber&&
-2\left[(\Phi-\Phi_o)-(\D_{\pp} v_s-\D_{\pp}v_o)\right]
\left[\chi(\D_{\pp}v'-\D_{\pp}^2v +2\Phi')\right]\bigg]
  -\frac{4}{\chi_s}\left(\Phi_s-2\Phi_o\right)\Ic\Phi
+4\left[\left(\frac{\HH'}{\HH^2}+1\right)\Phi
\right.\\ \nonumber&&\left.
-\frac{\HH'}{\HH^2}\Phi_o
-\left(\D_{\pp}v_s-\D_{\pp}v_o\right)
+\chi(\Phi'-\D_{\pp}\Phi)\right]\Ic\Phi'
-2\left[\left(\D_{\pp}v_s-\D_{\pp}v_o\right)
-(\Phi-\Phi_o)\right]
\\ \nonumber&&
\times\left[\frac{2}{\chi}\Ic\Phi+\Ic\frac{(\chi-\chi_s)\chi}{\chi_s}\nabla^2_{\o}\Phi
-2\Ic\Phi'-\Ic\chi\nabla_{\o}^2\Phi\right]+4\left[\left(\frac{\HH'}{\HH^2}-1\right)\Ic\Phi'-\Ic\chi\nabla^2_{\o}\Phi
\right.\\ \nonumber&&\left.
+\frac{2}{\chi_s}\Ic\Phi+\Ic\frac{(\chi-\chi_s)\chi}{\chi_s}\nabla^2_{\o}\Phi\right]\Ic\Phi'
-\left(1-\frac{1}{\chi_s\HH_s}\right)
\bigg(
4\left[\D_{\pp}v_o+2\D_{\pp}v_s+\left(-2\Phi+3\Phi_o\right)
\right.\\ \nonumber&&\left.
+\chi\left(\D_{\pp}v'_s-\D_{\pp}^2v_s+2\Phi'\right)\right]
\Ic\Phi'
-4\nabla_{\o} v_s\Ic\nabla_{\o}^i\Phi
-2\left[(\D_{\pp}v_s-\D_{\pp}v_o)-(\Phi-\Phi_o)\right]
\left[\Ic\chi\nabla^2_{\o}\Phi
\right.\\ \nonumber&&\left.
-\frac{4}{\chi}\Ic\Phi-\frac{2}{\chi}\Ic(\chi-\chi_s)\chi\nabla^2_{\o}\Phi\right]
+8\Ic\frac{\chi}{\chi_s}\nabla_{\o}^i\Phi'\Icpr\nabla_{\o}^i\Phi(\tilde{\chi})
+4\left[\Ic\chi\nabla^2_{\o}\Phi-\frac{2}{\chi_s}\Ic(\chi-\chi_s)\chi_s\nabla^2_{\o}\Phi
\right.\\ \nonumber&&\left.
-\frac{4}{\chi_s}\Ic\Phi+\Ic\Phi'\right]\Ic\Phi'
 \bigg)
-  4\Phi_o\left[2\Ic\frac{\chi}{\chi_s}\Phi'
  +2\Ic\frac{(\chi-\chi_s)}{\chi_s}
  \Phi'
  -2\Ic\frac{(\chi-\chi_s)\chi}{\chi_s}\Phi''\right]
  +16\Ic\frac{(\chi-\chi_s)}{\chi_s}\Phi\Phi'
   \\ \nonumber&&
  +\frac{2}{\chi_s}\I\Phi^2  
  - 2\left(6\Phi_o+\Phi_s\right)\Ic\frac{(\chi-\chi_s)\chi}{\chi_s}\nabla^2_{\o}\Phi
  -2\D_{\pp}v_o\left[\frac{4}{\chi_s}\Ic\Phi-\Ic\frac{(\chi-\chi_s)\chi}{\chi_s}\nabla_{\o}^2\Phi\right]
   \\ \nonumber&&  
  -4\Ic\frac{1}{\chi\chi_s}\Phi\Icpr\Phi(\tilde{\chi})
  +8\Ic\Phi'\Icpr\Phi'(\tilde{\chi})+8\Ic(\Phi\Phi')
  -4\Ic\frac{(\chi-\chi_s)}{\chi\chi_s}\left[\frac{1}{\chi}\Phi\Icpr\Phi(\tilde{\chi})+\Phi^2-2\Phi'\Ic\Phi(\tilde{\chi})\right]
  \\ \nonumber&&
  -4\Ic\frac{1}{\chi}\Phi\Icpr\frac{(\tilde{\chi}-\chi)\tilde{\chi}}{\chi}\nabla_{\o}^2\Phi(\tilde{\chi})+14\Ic\frac{(\chi-\chi_s)\chi}{\chi_s}\Phi\nabla_{\o}^2\Phi
  -4\Ic\frac{(\chi-\chi_s)}{\chi^2\chi_s}\Phi\Icpr{(\tilde{\chi}-\chi)\tilde{\chi}}\nabla_{\o}^2\Phi(\tilde{\chi})
  \\ \nonumber&&
  +4\Ic\frac{(\chi-\chi_s)^2}{\chi_s}\Phi\nabla^2_{\o}\Phi
  -4\Ic\Phi\Icpr\frac{\tilde{\chi}}{\chi}\nabla_{\o}^2\Phi(\tilde{\chi})
  +4\Ic\frac{(\chi-\chi_s)}{\chi_s}\nabla_{\o}^2\Phi\Ic\Phi
  +4\Ic\frac{(\chi-\chi_s)}{\chi_s}\Phi'\Ic\chi\nabla_{\o}^2\Phi
  \\ \nonumber&&
  +4\Ic\frac{(\chi-\chi_s)}{\chi}\Phi\Icpr\frac{(\tilde{\chi}-\chi)\tilde{\chi}}{\chi_s}\nabla_{\o}^2\Phi(\tilde{\chi})
   -8\Ic\frac{(\chi-\chi_s)}{\chi_s}\nabla_{\o i}\Phi\Icpr\frac{\tilde{\chi}}{\chi}\nabla_{\o}^i\Phi(\tilde{\chi})
  -8\Ic\frac{\chi}{\chi_s}\nabla_{\o i}\Phi\Icpr\nabla^i_{\o}\Phi(\tilde{\chi})
     \\ \nonumber&&
  -\frac{8}{\chi_s}\Ic\Icpr\nabla_{\o i}\Phi(\tilde{\chi})\Icpr\nabla_{\o}^i\Phi(\tilde{\chi})
  +12\Ic\frac{(\chi-\chi_s)\chi}{\chi_s}\nabla_{\o i}\Phi\nabla_{\o}^i\Phi
  -8\Ic\frac{(\chi-\chi_s)\chi}{\chi_s}\nabla_{\o i}\Phi'\Icpr\nabla_{\o}^i\Phi(\tilde{\chi})
  \\ \nonumber&&
  -2\Ic\frac{(\chi-\chi_s)}{\chi_s}\nabla_{\o}^2\Phi\Icpr(\tilde{\chi}-\chi)\tilde{\chi}\nabla_{\o}^2\Phi(\tilde{\chi})
  +8\Ic\frac{(\chi-\chi_s)}{\chi_s}\Icpr\nabla_{\o i}\Phi(\tilde{\chi})\Icpr\frac{{\tilde{\chi}}^2}{\chi}\nabla_{\o}^i\nabla_\o^2\Phi(\tilde{\chi})
  \\ &&
  -4\Ic\frac{(\chi-\chi_s)\chi}{\chi_s}\Icpr\nabla_{\o\<i}\nabla_{\o j\>}\Phi(\tilde{\chi})\Icpr\nabla_{\o}^{\<i}\nabla_{\o}^{ j\>}\Phi(\tilde{\chi})
  -4\Ic\frac{(\chi-\chi_s)}{\chi_s}\nabla_{\o i}\Phi\Icpr\frac{(\tilde{\chi}-\chi){\tilde{\chi}}^{2}}{\chi}\nabla_{\o}^i\nabla_{\o}^2\Phi(\tilde{\chi})\bigg\}
 \bigg\}\,.\label{bigboy}
 \end{eqnarray}
We have arranged this according to the manner in which we have derived the results:  first-order, then second-order induced scalar, induced vector, induced tensor and the quadratic part. The integrated parts are presented to simplify evaluation for practical purposes. An alternative presentation is given in Appendix B.

\subsection{Relation to the weak lensing magnification}

The magnification, $\mu$,  is related to the determinant of the amplification matrix, ${\bf{A}}$, according to $\mu= 1/\text{det}\, {\bf{A}}$~\cite{Schneider:2005ka}. In terms of the weak lensing shear $\gamma$  and convergence $\kappa$, the magnification is given by  $\mu=\left[(1-\kappa)^2-|\gamma|^2\right]^{-1}$. The determinant of the amplification matrix is proportional to the square of the area distance, $ \text{det}\, {\bf{A}}\propto D_{A}^2$, hence the magnification in terms of the observed redshift is given by
 \begin{eqnarray}
\mu (z_s)&=&\frac{(1+\hat{z}_s)^2}{\chi_s^2}\left\{1-2\left[\frac{\delta D_A}{D_A}+\left(1- \frac{1}{\HH_s (\nu_o-\nu_s)}\right)\delta z\right]
 -\left[\frac{ \delta^2 D_A}{D_A} +\left(1- \frac{1}{\HH_s\chi_s}\right)\delta^2 z
 + 2\bigg(\frac{\delta D_A}{D_A}
 +\frac{\d_{\nu}\delta z}{\HH_s}\bigg)
  \right. \right. \nonumber\\ &&\left.\left. \times
   \left(1- \frac{1}{\HH_s\chi_s}\right)\delta z+2\d_{\nu}\frac{\delta D_A}{\HH_sD_A} \delta z
 +  \left(
 \frac{\HH'_s}{\HH^2}- 1\right)\frac{(\delta z)^2}{\HH(\nu_s-\nu_o)}+6\left(\frac{\delta D_A}{D_A}+\left(1-\frac{1}{\chi_s\HH_s}\right)\delta z\right)^2\right]
 \right\}.\label{DAgen2}
 \end{eqnarray}

\section{Conclusion}

We have presented the full expression for the observed area distance-redshift relation to second-order on an arbitrary flat FLRW background~\eqref{bigboy}. Other than our assumption that $\Phi=\Psi$ at first-order, our formula is valid for any dark energy model~-- excluding only dark energy models with anisotropic stress. 
 We have not assumed explicitly that over-densities must be small. Consequently, our result should hold into the mildly nonlinear regime. 

We have presented the result in a ready-to-use form. Given a first-order potential $\Phi$, and a matter model, the second-order potential and velocities are fixed through the Einstein field equations~-- see e.g., \eqref{phi2}-- \eqref{delta2} in the case of LCDM. Each term in~\eqref{bigboy} is then directly computable. This can be implemented in N-body codes to correct for line-of-sight effects. Alternatively, it can be used to compute the spherical harmonic expansion of $\langle\hat D_A(\hat z_s)\rangle$, which then gives the deviation from the background geometry expected in the standard model. This latter `backreaction' effect has been postulated to be both large and small, and is very important to calculate accurately, in order to correctly determine the background geometry and parameters (see e.g., \cite{Rasanen:2011bm,Clarkson:2011zq,Clarkson:2011uk,Marozzi:2012ib,BenDayan:2012ct}).

\subsection*{ Acknowledgments}

We thank Camille Bonvin and Ruth Durrer for useful discussions and  Giovanni Marozzi and Gabriele Veneziano for comments. Most of the computations here were done with the help of the tensor algebra packages xAct /xPert~\cite{Brizuela:2008ra} and  xPand~\cite{Pitrou:2013hga}.
OU and RM are supported by the South African Square Kilometre Array
(SKA) Project, CC and RM are supported by the National Research Foundation
(South Africa), RM is supported by the Science \& Technology Facilities Council (UK) (grant no.
ST/K00090X/1), and all authors were supported by a Royal Society
(UK)/ NRF (SA) exchange grant.

\appendix

\section{Notation -- derivatives and integrals}
\label{Notation}


In the Minkowski background, the spatial derivative is decomposed into parts along the null geodesic and in the screen space: 
\be
\D_i=n_i\D\p+\nabla_{\o i}=n_i(\partial_\eta-\d_\lambda)+\nabla_{\o i}\,.
\ee
Then 
\begin{eqnarray}\label{eq:spatialdecomposition1}
\D_i\D_j  &=& \D_j \left[n_i\nabla\p   +  \nabla_{\bot i}\right]
=n_in_j\nabla\p ^2+ 2 n_{(i}\nabla_{\bot j)}\nabla\p  
+\frac{1}{\chi}\left(\gamma_{ij}-n_in_j\right)\nabla\p  
+ \nabla_{\bot i}\nabla_{\bot j}\,,
\end{eqnarray}
 where  we used 
  \begin{eqnarray}\label{eq:nexpansion}
  \D_i n_j&=& \frac{1}{(\lambda_o-\lambda)}\left(\gamma_{ij}-n_in_j\right)
  =\frac{1}{\chi}\left(\gamma_{ij}-n_in_j\right)\,.
  \end{eqnarray}
We also have 
\begin{eqnarray}
n^in^j\D_{i}\D_{j} X = \D_{\pp}^2 X = \frac{\d^2 X}{\d\lambda^2}-2 \frac{\d X'}{\d\lambda}+X''\,,
\end{eqnarray}
and the 3D Laplacian on the Minkowski background becomes 
\be
\D^2=\D\p^2+2\chi^{-1}\D\p+\nabla_\o^2.
\ee
For all derivatives, we assume for convenience that they act on the following term \emph{only}, e.g.  $\nabla_i XY=(\nabla_i X)Y$.
For integrals of $X$ down the past light cone \emph{from the observer}, we introduce the shorthand notation:
\be
\I X=\left[\I X\right](\lambda)=\int_{\lambda_o}^{\lambda} X(\tilde{\lambda})\d\tilde{\lambda}~~~\text{and}~~~\I XY=\left[\int_{\lambda_o}^{\lambda} X(\tilde{\lambda})\d\tilde{\lambda}\right]Y\,.
\ee
In this definition we have used the same convention that the operator $\I$ acts on the quantity immediately to the right of it, as for derivatives. The exception to this is that functions of $\lambda$ are assumed to be part of the integrand that is operated upon, e.g., $\I(\lambda_s-\lambda)X = \int_{\lambda_o}^{\lambda} (\lambda_s-\tilde{\lambda})X(\tilde{\lambda})\d\tilde{\lambda}$. Where there is a chance of confusion, integrals will be written out in full. For double integrals, 
\be
\I\Ipr X(\tilde{\lambda}) = \int_{\lambda_o}^{\lambda_s}\d\lambda\int_{\lambda_o}^{\lambda}\d\tilde{\lambda} X(\tilde{\lambda})\,.
\ee
The same notational conventions apply to integrals over $\chi$: 
\be
\Ic X=\int_0^{\chi_s}X\d\chi\,.
\ee

For derivatives of integrals, the identity
\begin{eqnarray}
\frac{\d}{\d x}\int_{a(x)}^{b(x)}f(x,t)\d t=f\left(x,b(x)\right)b'(x)-f\left(x,a(x)\right)a'(x)+\int^{b(x)}_{a(x)} {\partial f(x,t)\over \partial x}\d t\,,
\end{eqnarray}
leads to 
\begin{eqnarray}
\nabla_\|\int_{\lambda_o}^{\lambda_s}f(\eta,{\x})\d \lambda&=&f(\eta,{\x})\Big|_{\lambda_s}\frac{\d\lambda}{\d \chi}=-f(\eta,{\x})\Big|_{\lambda_s}\,,
\end{eqnarray}
using $\lambda = \lambda(\eta,\chi)$ 
and  $\d \lambda = \d \eta = -\d \chi$ along a light ray, and similarly for $\partial_\eta$.
For angular derivatives,  
\begin{equation}
\nabla_{\o i} \int^{\lambda_s}_{\lambda_o}f(\lambda, {\n})\d\lambda=\int^{\lambda_s}_{\lambda_o}\frac{(\lambda_o-\lambda)}{(\lambda_o-\lambda_s)}\nabla_{\o i}f(\lambda, {\n})\d\lambda\,.
\end{equation}





We can simplify double integrated integrals:
\begin{eqnarray}
\I \Ipr X  = \I(\lambda_s-\lambda) X\,,~~~\mbox{i.e.,}~~~
\int_{\lambda_o}^{\lambda_s}\d\lambda\int_{\lambda_o}^{\lambda}X(\tilde{\lambda})\d\tilde{\lambda}=\int_{\lambda_o}^{\lambda_s}(\lambda_s-\lambda) X(\lambda)\d\lambda\,.
\end{eqnarray}
Using integration by parts we have
\begin{eqnarray}
\I(\lambda_s-\lambda){\d_{\lambda} X } &=&(\lambda_o-\lambda_s) X_o + \I X   \,,\\
\I\left(\lambda_s-\lambda\right)\left(\lambda_o-\lambda\right){\d_\lambda}X
&=&\I (\lambda_o-\lambda)X+\I (\lambda_s-\lambda)X=\I(\lambda_s+\lambda_o-2\lambda)X\,,\\
\I\left(\lambda_s-\lambda\right)\left(\lambda_o-\lambda\right) {\d^2_\lambda X}&=&(\lambda_o-\lambda_s)(X_s+X_o)
+2\I X \,.
\end{eqnarray}

To differentiate integrals, the following is useful:
\bea
\d_\lambda\I(\lambda_s-\lambda)X=\I X\,,~~~\mbox{i.e.,}~~~
\frac{\d}{\d\lambda_s}\int_{\lambda_o}^{\lambda_s}\left(\lambda_s-\lambda\right)A(\lambda)\d \lambda=\int_{\lambda_o}^{\lambda_s}A(\lambda)\d \lambda\,.
\eea





\section{Notation -- screen space and scalar-vector-tensor decompositions}
\label{Notation2}

Consider a spatial vector $V^a$ (i.e., orthogonal to $u^a$),  on perturbed Minkowski spacetime. There are two decompositions: into scalar and vector modes,
\be
V_i=\nabla_i v\S+v_i\V,~~~\nabla_iv^i\Vd=0,
\ee
and a screen-space decomposition,
\be
V_i=n_i V\p+V_{\o i}\,,~~~V\p=n^iV_i,~ V_{\o i}=N_j^{~i}V_j.
\ee
Combining these,
\be
V^i=n^i(\D\p v\S+v\p\V)+\nabla_\o^i v\S+{v_\o\V}^i\,.
\ee
We can drop the S and V superscripts because there is no notational confusion, i.e. $\D\p v=\D\p v\S$, $v\p=v\p\V$ and $v_\o^i={v_\o\V}^i$. Consequently,
\be
V\p=\D\p v+v\p,~~~\text{and}~~~V_{\o}^i=\nabla_{\o}^i v+v_{\o}^i\,.
\ee

For a symmetric trace-free spatial tensor $W_{ab}$, the SVT decomposition is
\be
W_{ij}=\left(\D_i\D_j-\frac{1}{3}\delta_{ij}\D^2\right)w\S+2\D_{(i}w_{j)}\V+w_{ij}\T,~~~\D^iw_i=\D^iw_{ij}=0=w_i{}^i,
\ee
and the screen space decomposition is
\be\label{3t2dec}
W_{ij}=W\p \left(n_in_j-\frac{1}{2}N_{ij}\right)+2W_{\po(i}n_{j)}+ W_{\bot ij}\,,~~~W_{\po}^in_{i}=0=n_{i}W_\o^{ij}=N_{ij}W_\o^{ij}.
\ee
We use $\po$ to denote the part of a symmetric trace-free spatial tensor which has mixed components, i.e. along $n^a$ and in the screen space. 
Consequently,
\bea
W\p&=&\frac{1}{3}\left(2\D\p^2-\nabla_\o^2\right)w\S+2\D\p w\p\V+w\p\T \,,\\
W_{\po}^i&=&\D\p\nabla_{\o}^iw\S+\D\p {w_\o\V}^i+\nabla_\o^iw\p\V+{w_\po\T}^i\,,\\
W\p^{ij}&=&\nabla_\o^i\nabla_\o^jw\S-\frac{1}{2}N^{ij}(\D\p^2+\nabla_\o^2)w\S
+2\nabla_\o^{\<i}{w_\o\V}^{j\>}+{w_\o\T}^{ij}+\frac{1}{2}N^{ij}w\p\T\,.
\eea
In general, we should keep the V and T indicators on the symbols, unlike for the vector case, but we only have one tensor degree of freedom $h_{ij}$ in our analysis, so we drop the tensor indicator. 



\section{Second-order perturbation in the concordance model}
\label{concordancemodel}


For ease of reference we give the standard results for second-order perturbations for a flat LCDM background: we have $\mathcal{H}=aH_0\sqrt{\Omega_ma^{-3}+1-\Omega_m}$, and the first-order potential is $\Phi=g(\eta)\Phi_0(\bm x)$, where the growth suppression factor $g(\eta)$ is determined from $g''+ 3 \mathcal{H} g' +  a^2\Lambda g= 0$, with initial conditions at the end of the radiation era giving $\Phi_0(\bm x)$, such that $g_0=1$. The variables $\Phi\two,\Psi\two,\omega_i,h_{ij}$ are second order, and are sourced by terms $\mathcal{O}(\Phi^2)$. The full solutions to the second-order potentials may be found in~\cite{Bartolo:2005kv}; on small scales they are given by
\be\label{phi2}
\Phi\two=\Psi\two\simeq B_3(\eta) \nabla^{-2} \D_i\D^j(\D^i \Phi_0 \D_j \Phi_0 )+B_4(\eta) \D^i \Phi_0 \D _i\Phi_0,
\ee
where $B_{3,4}$ 
are functions of conformal time only, 
and consist of integrals over the growth suppression factor. On large scales they are no longer equal (see \cite{Bartolo:2005kv}). The second-order vectors are determined from~\cite{Lu:2008ju}
 \be \label{new solution for S}
\omega_i={16\over 3{\cal H}^2\Omega_m}\nabla^{-2} \Big\{
\nabla^2\Phi\,\D_i \left( \Phi'+{\cal H}\Phi\right)\Big\}^V\,,
 \ee
where $V$ denotes the vector contribution of the part in braces. These modes peak in power at the equality scale, and have the same spectrum as $\Phi$ below this scale, but the potentials have $\lesssim$1\% of the amplitude~\cite{Lu:2008ju}. Finally, the tensors are given by
\bea
&& h_{ij}'' + 2{\cal H} h_{ij}'
-\nabla^{2} h_{ij}  =  \Big\{-16\Phi\D_i\D_j\Phi
-8\D_i\Phi\D_j\Phi
 +
\frac{4}{\mathcal{H}^2\Omega_m}
\left[ {\cal H}^{2}\D_i\Phi\D_j\Phi +2\mathcal{H}\D_i\Phi\D_j\Phi' +
\D_i\Phi'\D_j\Phi' \right]\Big\}^T
\eea
where $T$ denotes a tensor projection~\cite{Ananda:2006af}. The induced gravitational wave background also peaks in power around the equality scale, and is surprisingly larger than the primordial background on these scales~\cite{Baumann:2007zm}.\\
Now, considering $\hat u^a$ as the geodesic CDM 4-velocity, the first-order velocity potential is given by
\be
v=-\frac{2}{3\mathcal{H}^2\Omega_m} \left(\HH \Phi + \Phi'\right)\,.
\ee
The second-order velocity has scalar $v\two$ and vector $v\two_i$ contributions, given by
\begin{eqnarray}
\HH^2\Omega_m\left[ v\two_i+ \D_iv\two\right] &=&  [\frac{1}{6} \nabla^2-\HH^2\Omega_m]\omega_i  -\frac{2}{3}\D_i[\HH\Phi\two
    +{\Psi\two}']
   +
 \frac{4}{3\Omega_m}\left[(\Omega_m-2)(\HH \Phi\D_i\Phi+\Phi'\D_i\Phi)-(2+3\Omega_m)\Phi\D_i\Phi' \right]  \nonumber\\&&
  -\frac{8}{9}\frac{1}{\HH^2\Omega_m}\left(3\HH \Phi'\D_i\Phi'-\nabla^2\Phi\D_i\Phi'-\HH\nabla^2\Phi\D_i\Phi\right)\,,\label{v2}
  \end{eqnarray}
which can be split into scalar and vector parts using a suitable projection in Fourier space. In real space this can be achieved operating first with $\D^{-2}\D^i$ to isolate the scalar part, then substituting the result to leave the vector.
The gauge-invariant density perturbation is given by 
\be
\delta=\frac{\delta\rho}{ \rho} =\frac{2}{a^2\rho}\left[\nabla^2\Phi-3\HH\left(\Phi'+\HH\Phi\right)\right]\,.
\ee
The second-order density perturbation is given by
\begin{eqnarray}
&&a^2 \delta^2\rho =  2\nabla^2\Psi\two-6 \left(\HH^2\Phi\two+\HH\Psi\two\right)+ 16\Phi\nabla^2\Phi
  + 6\D_k\Phi\D^k\Phi
+6\Phi'^2+ 24\HH^2 \Phi^2-6\mathcal{H}^2\Omega_m \D_kv\D^k v \,.\label{delta2}
\end{eqnarray}
We do not use these second-order solutions to the field and conservation in our expression for the area distance or redshift in what follows, preferring to write everything in terms of the metric and velocity potentials. If desired one could use the equations above to substitute for the velocity potential at first and second-order, and for the second-order metric potentials, in order to write $\hat D_A(\hat z_s)$ purely in terms of the first-order potential $\Phi_0$, with time dependence $g(\eta)$. Alternatively one may prefer to eliminate $\Phi\two$ in favour of $\delta\two\rho$, and so on. 


\section{Ricci Tensor}

The Ricci tensor is required in the Sachs equations. We present it here to second-order on a Minkowski background.
Here, $\DD$ is the spatial derivative associated with the spatial metric on the hyper-surface orthogonal to $u^a$.

\begin{description}
\item [\tt{First-order perturbation}]
\begin{eqnarray}
\delta R_{ab}&=&- 4 u_{(a}\DD_{b)}\Phi' + u_au_b\left(3 \Phi''+\DD_d\DD^d\Phi\right) + + \gamma_{ab}\left(\Phi''+ \DD_d\DD^d\Phi\right)\,,\nonumber\\
&=&
4n_{(a}u_{b)}\left(\frac{\d\Phi'}{\d\lambda}-\Phi''\right)-4u_{(a}\nabla_{\o b)}\Phi'
+\bigg[u_{a}u_{b}+n_{a}n_{b}+N_{ab}\bigg]
\nonumber\\ &&
\times\left[\frac{\d^2\Phi}{\d\lambda^2}-\frac{2}{\chi}\left(\frac{\d\Phi}{\d\lambda}-\Phi'\right)-2\frac{\d\Phi'}{\d\lambda}+4\Phi''+\nabla_{\o}^2\Phi\right]\,.
\end{eqnarray}

\item [\tt{Second-order scalar perturbation}]
(Here, $\Phi$ and $\Psi$ are the second-order potentials.)
\begin{eqnarray}
\delta^2\Sd R_{ab}&=& u_{a}u_{b}\left(3\Psi'+\DD_c\DD^c\Phi\right)-4u_{(a}\DD_{b)}\Psi'+h_{ab}\left(-\Psi''+\DD_c\DD^c\Psi\right)+
\DD_a\DD_b\Psi-\DD_a\DD_b\Phi\,,\nonumber\\ 
&=&u_{a}u_{b}\left[\frac{\d^2\Phi}{\d\lambda^2}
-\frac{2}{\chi}\left(\frac{\d\Phi}{\d\lambda}-\Phi'\right)-2\frac{\d\Phi'}{\d\lambda}+\Phi''+3\Psi''+\nabla_{\o }^2\Phi\right]-4u_{(a}\nabla_{\o b)}\Psi' \nonumber\\ &&
+4n_{(a}u_{b)}\left[\frac{\d\Psi'}{\d\lambda}-\Psi''\right]
+n_{a}n_{b}\left[-\frac{\d^2(\Phi-2\Psi)}{\d\lambda^2}-\frac{2}{\chi}\left(\frac{\d \Psi}{\d\lambda}-\Psi'\right)
+2\left(\frac{\d\Phi'}{\d\lambda}-\frac{2\d\Psi'}{\d\lambda}\right)
\right.  
\nonumber\\ && \left.
+\left(\Psi''-\Phi''\right)+\nabla_{\o}^2\Psi\right]
+n_{(a}\left[\nabla_{\o b}\frac{\d(\Phi-\Psi)}{\d\lambda}-\nabla_{\o b}(\Phi-\Psi)\right]+\nabla_{\o a}\nabla_{\o b}(\Psi-\Phi) \nonumber\\&&
+N_{ab}\left[\frac{\d^2\Psi}{\d\lambda^2}+\frac{1}{\chi}\left(\frac{\d\Phi}{\d\lambda}-\Phi'\right)-\frac{3}{\chi}\left(\frac{\d\Psi}{\d\lambda}-\Psi'\right)
-2\frac{\d\Psi'}{\d\lambda}+\nabla_{\o}^2\Psi\right]
\,.
\end{eqnarray}

\item [\tt{Second-order vector perturbation}]
\begin{eqnarray}\label{eq:Riccifirst}
\delta^2\Vd R_{ab}&=& -\left(\DD_{(a}\omega'_{b)}-u_{(a}\DD_c\DD^c\omega_{b)}\right)\,,\nonumber\\
&=&2u_{(a}\bigg[\frac{1}{\chi^2}\omega_{\o b)}-\frac{1}{\chi}
\left(\frac{\d \omega_{\o b)}}{\d\lambda}-\omega'_{\o b)}\right)
+\frac{1}{2}\left(\frac{\d^2\omega_{\o b)}}{\d\lambda^2}-\omega''_{\o b)}+\nabla_{\o}^2\omega_{\o b)}\right)-\frac{\d\omega'_{\o b)}}{\d\lambda}+\frac{1}{\chi}\nabla_{\o b)}\omega_{\pp}\bigg]\nonumber\\&&
+2n_{(a}u_{b)}\bigg[-\frac{1}{\chi^2}\omega_{\pp}-\frac{1}{\chi}\left(\frac{\d\omega_{\pp}}{\d\lambda}-\omega'_{\pp}\right)+\frac{1}{2}\left(\frac{\d^2\omega_{\pp}}{\d\lambda^2}+\omega''_{\pp}+\nabla_{\o}^2\omega_{\pp}\right)-\frac{\d\omega'_{\pp}}{\d\lambda}+\frac{1}{\chi}\nabla_{\o d}\omega_{\o}^b\bigg]\nonumber\\&&
+n_{(a}\bigg[-\frac{1}{\chi}\omega'_{\o b)}+\frac{\d\omega'_{\o b)}}{\d\lambda}-\omega''_{\o b)}-\nabla_{\o b)}\omega'_{\pp}\bigg]+n_{a}n_{b}\left[\frac{\d\omega'_{\pp}}{\d\lambda}-\omega''_{\pp}\right]-\frac{1}{\chi}N_{ab}\omega'_{\pp}-\nabla_{\o(a}\omega'_{\o b)}.
\end{eqnarray}

\item [\tt{Second-order tensor perturbation}]
\begin{eqnarray}
\delta^2\Td R_{ab} &=&\left( h_{ab}''-\DD_c\DD^c h_{ab}\right)\,,\nonumber\\
&=&-\frac{\d^2 h_{\o ab}}{\d\lambda^2}
+\frac{2}{\chi}\left(\frac{\d h_{\o ab}}{\d\lambda}-h'_{\o ab}\right)+2\frac{\d h'_{\o ab}}{\d \lambda}-\frac{4}{\chi}\nabla_{\o (a}h_{\po b)}-\nabla_{\o}^2 h_{\o ab}
 \nonumber\\&&
+\left(N_{ab}-2n_{a}n_{b}\right)\bigg[-\frac{3}{\chi^2}h_{\pp}-\frac{1}{\chi}\left(\frac{\d h_{\pp}}{\d\lambda}-h'_{\pp}\right)+\frac{1}{2}\left(\frac{\d^2 h_{\pp}}{\d\lambda^2}+\nabla_{\o}^2 h_{\pp}\right)-\frac{\d h'_{\pp}}{\d\lambda}\bigg]
\nonumber\\&&
n_{(a}\bigg[-\frac{2}{\chi^2}h_{\po b}
+\frac{2}{\chi}\left(\frac{\d h_{\po b}}{\d\lambda}-h'_{\po b}\right)
-\frac{\d^2 h_{\po b}}{\d\lambda^2}
+2\frac{\d h'_{\po b}}{\d\lambda}-\frac{3}{\chi}\nabla_{\po b}h_{\pp}
-\nabla_{\o}^2 h_{\po b}\bigg].
\end{eqnarray}

\item [\tt{Second-order quadratic terms}]
\begin{eqnarray}\label{eq:RicciNonlinear}
\delta^2\NLd R_{ab}&=& 4u_au_b\left[ 3\Phi\Phi''+ \Phi\DD_d\DD^d\Phi  - \DD_k\Phi\DD^k\Phi\right]- 8u_{(a}\left[\Phi'\DD_{b)}\Phi+\Phi\DD_{b)}\Phi'\right]
 + 4\DD_a\Phi\DD_b\Phi
 \nonumber\\&& 
 + 8 \Phi\DD_{a}\DD_b\Phi 
+ 4 h_{ab}\left[ \Phi'^2+ \Phi\Phi''+\Phi \DD_d\DD^d\Phi+ \DD_k\Phi\DD^k\Phi\right] \nonumber\\
&=&4u_{a}u_{b}\bigg[-\left(\frac{\d\Phi}{\d\lambda}\right)^2
+\Phi\frac{\d^2\Phi}{\d\lambda^2}
-\frac{2}{\chi}\left(\frac{\d\Phi}{\d\lambda}-\Phi'\right)
+2\Phi\left(\frac{\d\Phi'}{\d\lambda}+2\Phi''\right)
+2\Phi'\frac{\d\Phi}{\d\lambda}-(\Phi')^2+\Phi\nabla_{\o}^2\Phi
 \nonumber\\&&
-\nabla_{\o k}\Phi\nabla_{\o}^k\Phi\bigg]+8n_{(a}u_{b)}\bigg[2\Phi\frac{\d\Phi'}{\d\lambda}+\Phi'\left(\frac{\d\Phi}{\d\lambda}-\Phi'\right)-2\Phi\Phi''\bigg]
-8u_{(a}\bigg[\Phi'\nabla_{\o b}\Phi+2\Phi\nabla_{\o b}\Phi'\bigg] \nonumber\\&&
+8n_{(a}\bigg[-2\Phi\nabla_{\o b)}\frac{\d\Phi}{\d\lambda}-\left(\frac{\d\Phi}{\d\lambda}-\Phi'\right)\nabla_{\o b}\Phi+2\Phi\nabla_{\o b}\Phi'\bigg]
+4\nabla_{\o a}\Phi\nabla_{\o b}\Phi+8\Phi\nabla_{\o a}\nabla_{\o b}\Phi
 \nonumber\\&&
+4n_{(a}n_{b)}\bigg[2\left(\frac{\d\Phi}{\d\lambda}\right)^2+3\Phi\frac{\d^2\Phi}{\d\lambda^2}-\frac{2}{\chi}\left(\frac{\d\Phi}{\d\lambda}-\Phi'\right)\Phi
-6\Phi\frac{\d\Phi'}{\d\lambda}-4\left(4\frac{\d\Phi}{\d\lambda}-3\Phi'\right)\Phi'
 \nonumber\\&&
+4\Phi\Phi'+\Phi\nabla_{\o}^2\Phi+\nabla_{\o k}\Phi\nabla_{\o}^k\Phi\bigg].
\end{eqnarray}

\end{description}

\section{Alternative presentation of the area distance}
\label{alternative}
\small

This presentation groups terms into boundary terms and line of sight integrated terms.
\begin{eqnarray}
\hat D_A &=& a(\chi_s)\chi_s \bigg\{1+\left(1- \frac{1}{\HH_s \chi_s}\right)\Phi_o+\frac{1}{2}\bigg[\left(1- \frac{1}{\HH_s \chi_s}\right)\Phi\two_o-\left(1- \frac{1}{\HH_s \chi_s}\right)\omega_{\p o}-\left(\frac{\HH'}{\HH^2}-7\right)\Phi_o^2\bigg]
-\left(2- \frac{1}{\HH_s \chi_s}\right)\Phi_s \nonumber
\\ \nonumber&&
+\frac{1}{2}\bigg[- \Psi\two_s-\left(1- \frac{1}{\HH_s \chi_s}\right)\Phi_s+\frac{1}{2}\left(1- \frac{2}{\HH_s \chi_s}\right)\omega_{\p s}-\frac{1}{2} h_{\p s}
-\left(\frac{\HH'}{\HH^2}-\frac{2}{\chi_s\HH_s}-7\right)\Phi_s^2
-2\left(1-\frac{2}{\chi_s\HH_s}\right)\chi_s\Phi_s\Phi'_s
\\ \nonumber&&
-2\chi_s\Phi_s\D_{\pp}\Phi_s\bigg] 
+\frac{1}{\HH_s \chi_s}\D_{\pp}v_o
+\frac{1}{2}\bigg[\frac{1}{\HH_s \chi_s}\D\p  v_o\two-\frac{1}{\HH_s\chi_s}v_{\p o}-\left(\frac{\HH'}{\HH^2}-2\right)\D_{\pp}v_o\D_{\pp}v_o 
+\frac{1}{\chi_s\HH_s}\nabla_{\o i}v_o\nabla_{\o}^iv_o\bigg]
\\ \nonumber&&
+\left(1- \frac{1}{\HH_s \chi_s}\right)\D_{\pp}v_s
+\frac{1}{2}\bigg[\left(1- \frac{1}{\HH_s \chi_s}\right)\D\p  v\two_s+\left(1- \frac{1}{\HH_s \chi_s}\right)v_{\p s}
+\left(\frac{\HH'}{\HH^2}-\frac{2}{\chi_s\HH_s}-1\right)\D_{\pp}v_s\D_{\pp}v_s
\\ \nonumber&&
+2\chi\left(1-\frac{1}{\chi_s\HH_s}\right)\D_{\pp} v_s
\left(\D_{\pp}v'_s-\D_{\pp}^2v_s \right)
+\left(1-\frac{1}{\chi_s\HH_s}\right)\nabla_{\o i}v_s\nabla_{\o}^iv_s
-2\left(\frac{\HH'}{\HH^2}-\frac{1}{\chi_s\HH_s}+4\right)\Phi_s\Phi_o
\\ \nonumber&&
+2\left(1-\frac{2}{\chi_s\HH_s}\right)\chi_s\Phi_o\Phi'_s-2\chi\Phi_o\D_{\pp}\Phi_s
-2\left(\frac{\HH'}{\HH^2}-\frac{1}{\chi_s\HH_s}\right)\D_{\pp}v_s\D_{\pp} v_o-2\chi_s\left(1-\frac{1}{\chi_s\HH_s}\right)\D_{\pp}v_o\left(\D_{\pp}v'_s
-\D_{\pp}^2v_s \right)
\\ \nonumber&&
-2\left(\frac{\HH'}{\HH^2}+\frac{1}{2\chi_s\HH_s}+\frac{11}{2}\right)\Phi_o\D_{\pp}v_o
-2\left(\frac{\HH'}{\HH^2}-\frac{1}{\chi_s\HH_s}+3\right)\Phi_s\D_{\pp}v_s+2\left(\frac{\HH'}{\HH^2}-1\right)\Phi_o\D_{\pp}v_s+2\chi_s\left(1-\frac{2}{\chi_s\HH_s}\right)\Phi'_s\D_{\pp} v_s
\\ \nonumber&&
+2\chi_s\D_{\pp}\Phi_s\D_{\pp}v_s-2\chi_s\left(1-\frac{1}{\chi_s\HH_s}\right)\Phi_s(\D_{\pp}v'_s-\D_{\pp}^2v_s)\,
-2\chi_s\left(1-\frac{2}{\chi_s\HH_s}\right)\Phi'_s\D_{\pp}v_o
 -2\chi_s\D_{\pp}\Phi_s\D_{\pp}v_o
 \\ \nonumber&&
+2\left(\frac{\HH'}{\HH^2}+2\right)\Phi_s\D_{\pp}v_o
+2\left(1-\frac{1}{\chi_s\HH_s}\right)\chi_s\Phi_o(\D_{\pp}v'_s-\D_{\pp}^2v_s)\bigg] \,
+
\frac{2}{\chi_s}  \Ic \Phi
+ \frac{1}{2}\bigg[\frac{1}{\chi_s} \Ic(\Phi\two+ \Psi\two)
\\ \nonumber&&
-\frac{1}{\chi_s} \Ic\frac{(\chi-\chi_s)}{\chi} \omega_{\pp} 
+ \frac{1}{\chi_s}\Ic 3\frac{(\chi-\chi_s)}{\chi} h_{\pp} 
 \bigg]
-2\left(1- \frac{1}{\HH_s \chi}\right)\Ic  \Phi'+\frac{1}{2}\bigg[-\left(1- \frac{1}{\HH_s \chi_s}\right)\left(\Ic\left( \Phi\twod' + \Psi\twod'\right)\right)
 \\ \nonumber&&
-\Ic\left(1- \frac{1}{\HH_s \chi_s}+\frac{\left(2\chi-\chi_s\right)}{2\chi_s}\right){\omega\p\two}'
+\Ic \left(1- \frac{1}{\HH_s \chi_s}-\frac{\left(2\chi-\chi_s \right)}{\chi_s}\right){h\p}'\bigg]
+ \Ic \frac{{(\chi-\chi_s) \chi}}{\chi_s}\nabla^2_{\bot}\Phi 
\\ \nonumber&&
+\frac{1}{2}\bigg[\frac{1}{2}\Ic \frac{(\chi-\chi_s)\chi}{\chi_s} \nabla_{\bot}^2( \Phi\two+\Psi\two)-\frac{1}{2}\Ic\frac{ (\chi-\chi_s)\chi}{\chi_s}\nabla_{\bot}^2\omega_{\pp}- \frac{1}{2} \Ic\frac{(\chi-\chi_s)\chi}{\chi_s} \nabla_{\bot}^2 h_{\pp}\bigg]\,
+\frac{1}{2}\bigg\{-2\Phi_o\left[\left(1+\frac{2}{\chi_s\HH_s}\right)\frac{2}{\chi}\Ic\Phi
\right.\\ \nonumber&&\left.
+2\left(\frac{\HH'}{\HH^2}+\frac{3}{\chi_s\HH_s}-4\right)\Ic\Phi'
-\left(2+\frac{1}{\chi_s\HH_s}\right)\Ic\chi\nabla_{\o}^2\Phi
+\left(9+\frac{2}{\chi_s\HH_s}\right)\Ic\frac{(\chi-\chi_s)\chi}{\chi_s}\nabla^2_{\o}\Phi
+ 4\Ic\frac{\chi}{\chi_s}\Phi'
\right.\\ \nonumber&&\left.
  +4\Ic\frac{(\chi-\chi_s)}{\chi_s}
  \Phi'-4\Ic\frac{(\chi-\chi_s)\chi}{\chi_s}\Phi'' 
    \right]
+2\Phi_s\bigg[\left(1-\frac{1}{\chi_s\HH_s}\right)\frac{4}{\chi}\Ic\Phi+2\left(\frac{\HH'}{\HH^2}-\frac{2}{\chi_s\HH_s}+2\right)\Ic\Phi'
  \\ \nonumber&&
-\left(2+\frac{1}{\chi_s\HH_s}\right)\Ic\chi\nabla_{\o}^2\Phi
+2\left(1-\frac{1}{\chi_s\HH_s}\right)\Ic\frac{(\chi-\chi_s)\chi}{\chi_s}\nabla^2_{\o}\Phi\bigg]
+2\D_{\pp}v_o\bigg[\frac{2}{\chi}\left(1-\frac{2}{\chi_s\HH_s}\right)\Ic\Phi-2\Ic\Phi'
\\ \nonumber&&
-\left(2-\frac{1}{\chi_s\HH_s}\right)\Ic\chi\nabla_{\o}^2\Phi
+2\left(2-\frac{1}{\chi_s\HH_s}\right)\Ic\frac{(\chi-\chi_s)\chi}{\chi_s}\nabla^2_{\o}\Phi
\bigg]
-2\D_{\pp}v_s
\bigg[\frac{2}{\chi_s}\left(5-\frac{1}{\chi_s\HH_s}\right)\Ic\Phi-4\left(1-\frac{1}{\chi_s\HH_s}\right)\Is\Phi'
\\ \nonumber&&
-\left(2-\frac{1}{\chi_s\HH_s}\right)\Ic\chi\nabla_{\o}^2\Phi
+\left(3-\frac{2}{\chi_s\HH_s}\right)\Ic\frac{(\chi-\chi_s)\chi}{\chi_s}\nabla^2_{\o}\Phi
\bigg]
+4\left(1-\frac{1}{\chi_s\HH_s}\right)\nabla_{\o} v_s\Ic\nabla_{\o}^i\Phi
\\ \nonumber&&
+4\left[\frac{2}{\chi_s}\left(3-\frac{2}{\chi_s\HH_s}\right)\Ic\Phi+\left(\frac{\HH'}{\HH^2}+\frac{1}{\chi_s\HH_s}-2\right)\Ic\Phi'
-\left(2-\frac{1}{\chi_s\HH_s}\right)\Ic\chi\nabla^2_{\o}\Phi
+\left(3-\frac{2}{\chi_s\HH_s}\right)\Ic\frac{(\chi-\chi_s)\chi}{\chi_s}\nabla^2_{\o}\Phi\right]\Ic\Phi'
\\ \nonumber&&
-4\chi\left(1-\frac{2}{\chi_s\HH_s}\right)\chi_s\Phi'_s\Ic\Phi'
-4\chi_s\D_{\pp}\Phi_s\Ic\Phi'
-4\left(1-\frac{1}{\chi_s\HH_s}\right)\left[\chi\left(\D_{\pp}v'_s-\D_{\pp}^2v_s\right)\right]
\Ic\Phi'
+16\Ic\frac{(\chi-\chi_s)}{\chi_s}\Phi\Phi'
\\ \nonumber&&
+\frac{2}{\chi_s}\I\Phi^2
  -4\Ic\frac{1}{\chi\chi_s}\Phi\Icpr\Phi(\tilde{\chi})
  +8\Ic\Phi'\Icpr\Phi'(\tilde{\chi})+8\Ic\Phi\Phi'
  -4\Ic\frac{(\chi-\chi_s)}{\chi\chi_s}\left[\frac{1}{\chi}\Phi\Icpr\Phi(\tilde{\chi})+\Phi^2-2\Phi'\Icpr\Phi(\tilde{\chi})\right]
  \\ \nonumber&&
 -4\Ic\frac{1}{\chi}\Phi\Icpr\frac{(\tilde{\chi}-\chi)\tilde{\chi}}{\chi_s}\nabla_{\o}^2\Phi(\tilde{\chi})
  -4\Ic\frac{(\chi-\chi_s)}{\chi^2\chi_s}\Phi\Icpr{(\tilde{\chi}-\chi)\tilde{\chi}}\nabla_{\o}^2\Phi(\tilde{\chi})
  +4\Ic\frac{(\chi-\chi_s)^2}{\chi_s}\Phi\nabla^2_{\o}\Phi
    \\ \nonumber&&
  -4\Ic\Phi\Icpr\frac{\tilde{\chi}}{\chi}\nabla_{\o}^2\Phi(\tilde{\chi})
  +14\Ic\frac{(\chi-\chi_s)\chi}{\chi_s}\Phi\nabla_{\o}^2\Phi 
  +4\Ic\frac{(\chi-\chi_s)}{\chi_s}\nabla_{\o}^2\Phi\Icpr\Phi(\tilde{\chi})
    \\ \nonumber&&
  +4\Ic\frac{(\chi-\chi_s)}{\chi_s}\Phi'\Icpr\tilde{\chi}\nabla_{\o}^2\Phi(\tilde{\chi})
  +4\Ic\frac{(\chi-\chi_s)}{\chi}\Phi\Icpr\frac{(\tilde{\chi}-\chi)\tilde{\chi}}{\chi}\nabla_{\o}^2\Phi(\tilde{\chi})
 -8\Ic\frac{(\chi-\chi_s)}{\chi_s}\nabla_{\o i}\Phi\Icpr\frac{\tilde{\chi}}{\chi}\nabla_{\o}^i\Phi(\tilde{\chi})
   \\ \nonumber&& 
  -8\Ic\frac{\chi}{\chi_s}\nabla_{\o i}\Phi\Icpr\nabla^i_{\o}\Phi(\tilde{\chi}) 
  -\frac{8}{\chi_s}\Ic\Icpr\nabla_{\o i}\Phi(\tilde{\chi})\Icpr\nabla_{\o}^i\Phi(\tilde{\chi})\,,
  +12\Ic\frac{(\chi-\chi_s)\chi}{\chi_s}\nabla_{\o i}\Phi\nabla_{\o}^i\Phi
    \\ \nonumber&&
  -8\Ic\frac{(\chi-\chi_s)\chi}{\chi_s}\nabla_{\o i}\Phi'\Icpr\nabla_{\o}^i\Phi(\tilde{\chi})
  -8\left(1-\frac{1}{\chi_s\HH_s}\right)
\Icpr\frac{\tilde{\chi}}{\chi}\nabla_{\o}^i\Phi'(\tilde{\chi})\Icpr\nabla_{\o}^i\Phi(\tilde{\chi})
\\ \nonumber&&
 -4\Ic\frac{(\chi-\chi_s)\chi}{\chi_s}\Icpr\nabla_{\o\<i}\nabla_{\o j\>}\Phi(\tilde{\chi})\Icpr\nabla_{\o}^{\<i}\nabla_{\o}^{ j\>}\Phi(\tilde{\chi})
-2\Ic\frac{(\chi-\chi_s)}{\chi_s}\nabla_{\o}^2\Phi\Icpr(\tilde{\chi}-\chi)\tilde{\chi}\nabla_{\o}^2\Phi(\tilde{\chi})
    \\ &&
  +8\Ic\frac{(\chi-\chi_s)}{\chi_s}\Icpr\nabla_{\o i}\Phi(\tilde{\chi})\Icpr\frac{{\tilde{\chi}}^2}{\chi}\nabla_{\o}^i\nabla^2\Phi(\tilde{\chi})
  -4\Ic\frac{(\chi-\chi_s)}{\chi_s}\nabla_{\o i}\Phi\Icpr\frac{(\tilde{\chi}-\chi){\tilde{\chi}}^2}{\chi}\nabla_{\o}^i\nabla_{\o}^2\Phi(\tilde{\chi})
\bigg\}
 \bigg\}\,.
 \end{eqnarray}

\normalsize


\bibliographystyle{apsrev4-1}
 \bibliography{DistanceRef}

\end{document}